\documentclass[manuscript]{acmart}

\setcopyright{none}

\newcommand{\probe}{InfoVid\xspace}
\newcommand{\probes}{InfoVids\xspace}

\newcommand{\staticVis}{\textsc{Baseline}\xspace}
\newcommand{\injuryVis}{\textsc{InjuryVis}\xspace}
\newcommand{\walmartVis}{\textsc{WalmartVis}\xspace}
\newcommand{\napoleonVis}{\textsc{NapoleonVis}\xspace}
\newcommand{\minardVis}{\napoleonVis}
\newcommand{\airplaneVis}{\textsc{AirplaneVis}\xspace}

\newcommand{\squote}[1]{`#1'} 
\newcommand{\myquote}[1]{\emph{``#1''}} %
\newcommand{\pquote}[2]{\myquote{#1}~(#2)}
\newcommand{\mylongquote}[1]{           %
    \begin{quote}\emph{#1}\end{quote}}

\newcommand{\edits}[1]{\textcolor{black}{#1}}

\usepackage{booktabs} %
\usepackage{enumitem}
\usepackage{graphicx} %
\RequirePackage[l2tabu, orthodox]{nag} %
\usepackage{microtype} %
\usepackage[subtle]{savetrees} %
\usepackage{subcaption}
\usepackage{xspace} %
\usepackage{xcolor}

\graphicspath{{figures/}} %

\usepackage{epigraph}

\copyrightyear{2025}
\acmYear{2025}
\setcopyright{none}
\acmPrice{15.00}
\acmConference[]{}{}{}
\acmBooktitle{}

\acmPrice{}
\acmDOI{}
\acmISBN{}

\begin{document}
\title{\probes: ~Reimagining~the~Viewer~Experience with Alternative Visualization-Presenter Relationships}

\settopmatter{authorsperrow=5}

\author{Ji Won Chung}
\affiliation{\institution{Brown University}}
\author{Tongyu Zhou}
\affiliation{\institution{Brown University}}
\author{Ivy Chen}
\affiliation{\institution{Brown University}}
\author{Kevin Hsu}
\affiliation{\institution{Brown University}}
\author{Ryan A. Rossi}
\affiliation{\institution{Adobe Research}}
\author{Alexa Siu}
\affiliation{\institution{Adobe Research}}
\author{Shunan Guo}
\affiliation{\institution{Adobe Research}}
\author{Franck Dernoncourt}
\affiliation{\institution{Adobe Research}}
\author{James Tompkin}
\affiliation{\institution{Brown University}}
\author{Jeff Huang}
\affiliation{\institution{Brown University}}

\renewcommand{\shortauthors}{Ji Won Chung et al.}

\begin{abstract}
\edits{Traditional data presentations typically separate the presenter and visualization into two separate spaces--the 3D world and a 2D screen--enforcing visualization-centric stories. To create a more human-centric viewing experience, we establish a more equitable relationship between the visualization and the presenter through our \probes. These infographics-inspired informational videos are crafted to redefine relationships between the presenter and visualizations. As we design \probes, we explore how the use of layout, form, and interactions affects the viewer experience. We compare \probes against their baseline 2D `slides' equivalents across 9 metrics with 30 participants and provide practical, long-term insights from an autobiographical perspective. Our mixed methods analyses reveal that this paradigm reduced viewer attention splitting, shifted the focus from the visualization to the presenter, and led to more interactive, natural, and engaging full-body data performances for viewers. Ultimately, \probes helped viewers re-imagine traditional dynamics between the presenter and visualizations.}
\end{abstract}



\begin{teaserfigure}
\includegraphics[width=\textwidth]{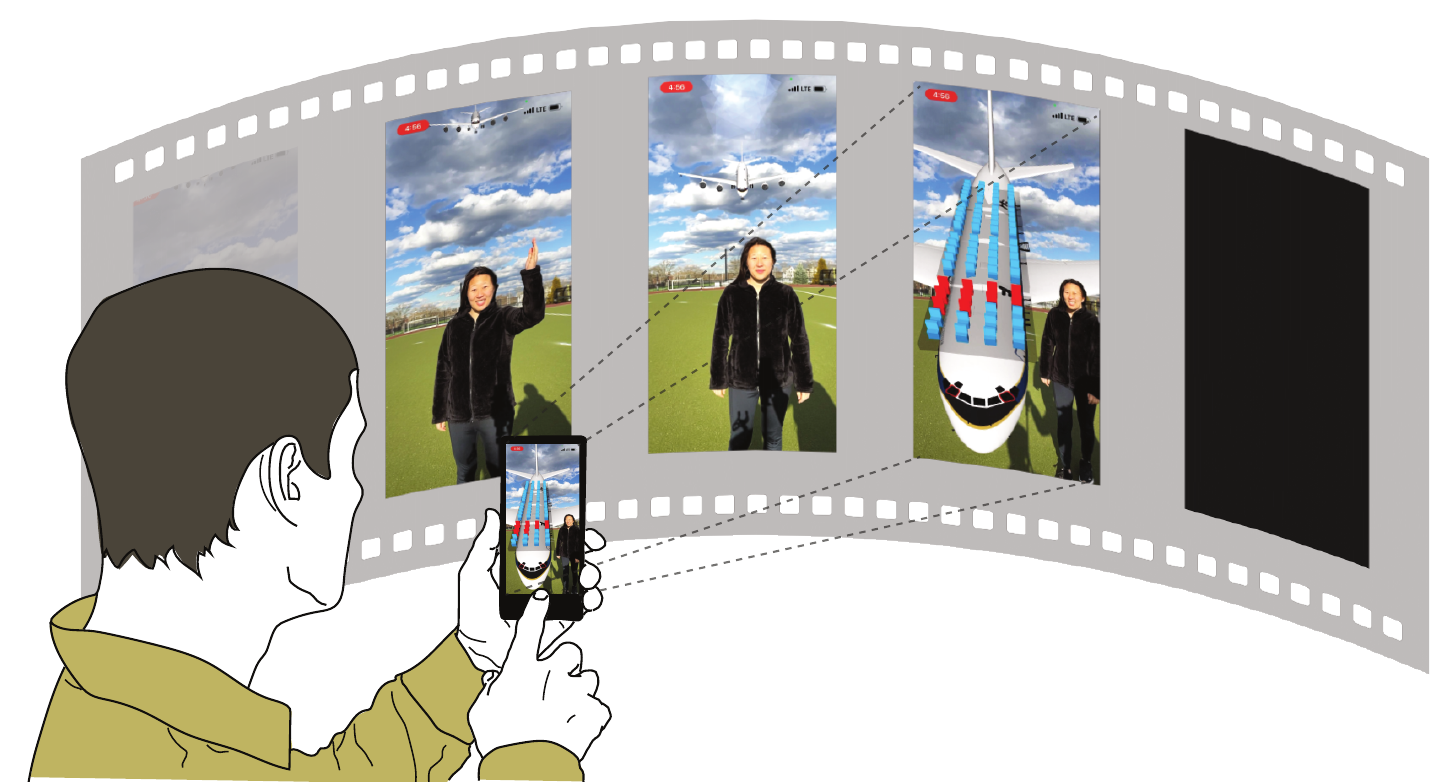}
  \caption[]{\textsc{\probes} provide viewers with a new visualization-presenter paradigm. By visualizing the presenter and the data within a shared 3D space, \probes redefine the relationship between them. \emph{Film Strip Left:} A virtual plane enters from the sky. \emph{Film Strip Center:} The background and clouds immerse the presenter within the frame and contextualize the plane. \emph{Film Strip Right:} The presenter's full-body next to the plane emphasizes their relative sizes. \textbf{The presenter is an external, hired actor.}}
  \Description[Teaser Figure]{Shows a drawn person holding up a smartphone and pointing to the phone screen. The phone screen is enlarged to show three images on top of a film roll to emphasize it is a moving image or a video. In the left image, a presenter is standing on a football field with her left arm pointing upward towards the sky. Above her head, a 3D rendered plane is approaching. In the middle image, the presenter stands in the same position and the plane gets closer to her. In the rightmost figure, the presenter is standing to the right of the image, letting the plane hover next to her. On top of the 3D plane, 3D models of seats appear. Most of them are blue seats, but some seats near the front of the plane are highlighted in red.}
  \label{fig:teaser}
  \vspace{2mm}
\end{teaserfigure}

\maketitle

\section{Introduction}
\epigraph{Now I'm going to try something I’ve never done before:\\ Animating the data in real space.}{\emph{Prof.~Hans Rosling}}

Data visualizations convey both the data and the ideas behind them. As Rosling demonstrates in his widely-viewed presentation \textit{200 Countries, 200 Years, 4 minutes}~\cite{hansRosling}, how you narrate the story behind the data can be just as important as the visualization itself---or even more so---in conveying the ideas. Rosling's body and the data \textit{co-exist} in the video space, letting him synchronize his body movements with animated 2D visualizations. His passion is evident as he kneels and gesticulates to demonstrate rises in life expectancy. 

Despite Rosling's captivating performance in 2011, \edits{modern visualization presentations by design typically separate presenters from the content to draw viewers' attention towards the visualizations. In many commercial, teleconferencing presentations (e.g. Zoom, Teams, Google Meet), presenters are compartmentalized away in a little box in the corner to maximize view of the visualizations. Such visualization-centric formats implicitly send the viewer a message to prioritize the visualization over the presenter. However, establishing an imbalanced relationship is a design choice and comes with trade-offs. The viewer misses the \textit{human element}, an opportunity to fully connect with the presenter as the presenter's body language and expressions are minimized. Imagine Rosling's performance restricted to a little box---would we, as viewers, feel the same level of passion and engagement?}

\edits{What if we offer viewers an alternative presentation paradigm that by design establishes an \textbf{equitable relationship} between the visualization and presenter? Rather than having the two compete for screen space, what if we integrate the visualization within the same space of the presenter, allowing the viewer to see both within the same frame? How would this shift impact the way \textit{viewers} engage with the content, the presenter, and the overall experience compared to more traditional formats? And in the process of designing and evaluating such presentations, what can we learn about designing for alternative visualization presentation systems?}

\edits{To answer these questions, we conduct an \textbf{exploratory investigation of the viewer experience with four custom-constructed case technology probes~\cite{hutchinson2003technology}, \probes}. These are informational videos inspired by `shorts' and infographics, iteratively designed over the span of four months. Each \probe showcases the viewer a unique spatial arrangement of the visualization and presenter, while ensuring the viewer has a full view of both at all times.} 

\edits{To understand the potential advantages and challenges of the \probe from the viewer's perspective, we ask 30 viewers in public to compare an \probe with its `baseline' equivalent. The latter includes the same content as the \probe, but adapts the format of a more traditional, videoconferencing presentation with visualizations. The baseline is not used to evaluate whether our probes are `better', but as a method to help viewers articulate the differences between the two formats, which may be difficult with the probe alone given that they are encountering a new medium. Viewers compare the two based on eight metrics that may influence the viewing experience: perceived presenter immersion, engagement, and co-presence with the visualization, natural body movement, enjoyability, storytelling, information understanding, and viewer's attention between the presentation and the visualization. We conduct a semi-structured interview to understand what elements in the \probe affected their viewing experience. Finally, we share the lessons learned from our viewer evaluations and the process of designing the \probes from an autobiographical perspective.}

\edits{Our exploratory investigation with \probes reveals valuable design implications for alternative, visualization presentation systems. First, we learn that the simple design condition of having both the presenter and the visualization in the same frame affects the viewing experience by compelling the presentation designer to be more conscious of the relationship between the visualization and the presenter --- an overlooked factor when the presenter is sidelined to a small box. In addition, while most viewers find \probes more enjoyable and pay more attention to the presenter over the visualization than the baseline, we discover how varying social expectations among the presenter, visualization, and viewer influence the viewing experience. Lastly, situating the presenter and the visualization in the same space enables viewers to consider the presenter's body as \textit{connected to the flow of the story}, an experience that is otherwise absent when the two are not co-located. 
}

\section{Related Work} \label{sec:related-work} 
\subsection{\edits{Different Ways to Communicate Visual Data}}
Data is more than just numbers---it tells a story, a message \textit{through} the numbers. To tell an effective story, we need the right tools. Different tools change \textit{how} we present information to the viewer. To interactively explore data stories, a viewer can use web-based visualizations such as D3.js~\cite{bostock2011d3}, Vega~\cite{satyanarayan2016vega}, or PortalInk~\cite{zhou2024portalink}. To unfold a story in real-time with the viewers, the presenter can draw a story using SketchStory~\cite{lee2013sketchstory}. To visually captivate viewers and convey a singular, simple message to them, we can create infographics~\cite{lo2022misinformed,li2014chart,lu2020exploring,zhou2024epigraphics}. 

\edits{More recently, as tools have evolved, a new paradigm for the viewing experience has emerged. Information now disseminates on social media platforms via short-form videos, or `shorts'. To engage viewers, `shorts' situate the presenter in the same space as the augmented virtual content, allow the presenter to interact directly with the virtual elements, and even use hard cuts and quick transitions to capture the viewer attention~\cite{wang2019comparing, wang2020humor, hassoun2023practicing, yang2023beyond}. This format stands apart from `data videos'~\cite{amini2015understanding,amini2018hooked,sallam2022towards}, as it places the presenter themselves at the core function of the video's narration and storytelling. In addition to the presenter's voice, their body language and facial expressions are central to the viewing experience. }

\edits{Such `shorts' aim to engage day-to-day viewers and inform them in a short amount of time. They often employ a more casual, and even playful narrative style and visualizations (e.g. memes, GIFs) and are not as formal as traditional presentations with visualizations~\cite{wang2020humor,zhu2020health}. They are like the infographic equivalent of videos; both use visual embellishments and focus on delivering simple and engaging data narratives to reach a wider audience~\cite{bateman2010useful, harrison2015infographic, li2014chart}. Inspired by such `shorts', we design new informational videos, or \textit{InfoVids}, that integrate presenters and visualizations in the same frame. 
} 

\edits{Question is, how do viewers respond to the integration of visualizations into `shorts' --- a new layout, different from traditional presentation styles? The literature has limited understanding on these new forms of presentations, and on what factors impact the viewer experience.} This is, in part, because AR data visualization tools have traditionally focused on using augmented reality to explore multidimensional scientific data~\cite{luo2021exploring, hubenschmid2021stream, yang2020tilt,satriadi2022tangible, tong1912exploring}, enhance human information processing capabilities~\cite{rajaram2022paper,chen2020augmenting, chen2023iarvis}, or enhance immersion with the data~\cite{cordeil2019iatk, sicat2018dxr, langner2021marvis, chen2019marvist}. \edits {They do not explore how the viewing experience is affected when AR is leveraged to convey information in more casual, presentation settings---this is the very gap in knowledge this paper seeks to fill. }

\subsection{\edits{Implicit Relationships Defined by Presenter-Visualization Interactions}}
\edits{To transform and enhance the viewing experience, previous works have explored innovative ways a presenter could interact with visual graphics. Both ChalkTalk~\cite{perlin2018chalktalk} and Saquib et al.~\cite{saquib2019interactive} allow the presenter to interact with 2D graphics and sketches. CLIO~\cite{davis2023multimodal} and Reality Talk~\cite{liao2022realitytalk} demonstrate a new viewing experience by allowing the presenter to interact with 2D visuals with their voice and hand gesture interactions. Hall et al. designs their own hand gesture language and experiments with different layouts of the visualization to ensure the scientific integrity of the visualizations for the viewers~\cite{hall2022augmented}.}

\edits{Other works investigate how to enhance and transform the viewing experience by coupling or binding visualizations with the presnter's body.} BodyVis engages the viewers with anatomy visualizations by using a wearable e-textile shirt~\cite{norooz2015bodyvis}. Within the realms of AR, MagicMirror~\cite{subramonyam2015sigchi} and mirracle~\cite{blum2012mirracle} both overlay medical or biodata visualizations over the presenter's body. RealitySketch uses AR to analyze different physical motions, including those of people~\cite{suzuki2020realitysketch}. \edits{In other cases, the coupling provides viewers with more to creative and artistic presentations.} HandAvatar enables the presenter to create non-humanoid puppet performances by binding virtual animals to the hand~\cite{jiang2023handavatar}. Pei et al.'s Hand Interfaces framework uses hands to generate an associated virtual object such as wands and healing potions~\cite{pei2022hand}. 

\edits{Such interactions not only transform the viewing experience, but more importantly, these interactions also implicitly define the relationship between the presenter and the visualization. Interactions establish who can affect what, and the parties involved in the interaction imply the nature of the relationship between them. For example, the presenter's ability to move around a visualization means the visualization is secondary, at will, to the presenter. Understanding the nature of the relationship is crucial because the relationship itself can convey a message to the viewers. However, this aspect is overlooked in previous works, and this paper aims to explore the factors that shape such relationships and understand how they may affect the viewing experience. }

\section{Designing \probes}
\edits{To inform the development of the tools needed to build and design four case \probes, we engaged in an iterative design process over the course of nine months. Each \probe serves as a technology probe, used to `find out about the unknown' and we adopt this approach to explore how we can `challenge pre-existing ideas and influenc[e] future'~\cite{hutchinson2003technology} visualization presentation technologies. This section outlines the development of our iterations and the lessons learned throughout the process. Because this design process required interdisciplinary knowledge of augmented reality, visualizations, and presentations---an intersectional expertise not commonly found among participants---we chose to draw our insights from an autobiographical perspective~\cite{neustaedter2012autobiographical,desjardins2018revealing,zhou2024portalink,huang2023irchiver} to offer a deeper understanding of the challenges and decisions involved in the process.} 

\subsection{\edits{Designing for an Equitable Relationship}}

\edits{Our primary aim was to design a format that establishes an equitable relationship with presenter and the visualization, from \textit{both} the presenter and the viewer's perspective. To achieve this, we first integrated the spatial dimensions of the visualizations with that of the presenter. Drawing from previous works~\cite{saquib2019interactive,liao2022realitytalk,hall2022augmented,davis2023multimodal,perlin2018chalktalk,hansRosling}, we overlaid a 2D visualization of bar charts and scatter plots onto the same screen of the presenter using augmented reality (ARKit on a iPhone). While, from the viewer's perspective, this setup placed the presenter at the center of the frame and resolved the visual divide between the presenter and the visualization (e.g. how the presenter is compartmentalized in a box in teleconferencing systems), we observed from our own experiences acting as presenter that this design still did not fully establish an equitable relationship.}

\subsubsection{Redefining Relationships Using the Third Dimension}
\edits{The flat 2D form of the visualization warranted that we, as the presenters, compromise our movements in 3D space for the visualizations. When we took a viewer's perspective, and reviewed recordings of ourselves presenting with this design, we frequently observed ourselves `miming'. As presenters, we were moving \textit{around} the virtual visualization and using awkward body language only in the horizontal dimension. We never utilized the third, $z$-dimension, as we might in a real-life presentation in 3D space. These awkward body movements are noticeable distractions, or `breaks in presence'~\cite{slater2000virtual, slater2003physiological}, that disrupt the viewer experience. In other words, integrating spaces was not enough to establish an equitable relationship; we were compromising bodily freedom and still prioritizing the visualization by design.}

\edits{To create an equitable relationship, we had to also ensure that the \textit{interactions} between the visualizations and presenter appeared balanced --- the presenter's body movements should not look compromised to the viewer. Following our lesson from the previous iteration, we introduced three-dimensionality to the form of the visualization. Our objective was to understand whether having the visualization occupy the same physical space as the presenter would reduce awkwardness in body movements and interactions. While the added third dimension allowed for more staging flexibility and enabled us, as the presenters, to move beyond the constraints a 2D plane, we still found ourselves walking and moving around the visualizations. The visualization continued to affect our interactions. However, we realized that this was not the crux of the issue.}

\subsubsection{Teeter the Balance with ``Wearable'' Visualizations}
\edits{The imbalance continued because the interaction was not bidirectional --- the visualization affected our interactions as presenters, but not vice versa. Thus, to equalize their relationship and send the message to the viewers that the two were of equal standing, we incorporated gestures that allowed the presenter to control the visualization, as informed by previous works~\cite{hall2022augmented,liao2022realitytalk,suzuki2020realitysketch,davis2023multimodal}. As a result, we now had two primary directions of interaction: the presenter could affect the visualization and vice versa.}

\edits{However, this process raised another question: could we design an even more equitable interaction, where control was not unidirectional, but mutual, or \textit{simultaneously} bidirectional? Referencing previous works on body-augmented visualizations~\cite{norooz2015bodyvis,blum2012mirracle,subramonyam2015sigchi,jiang2023handavatar,pei2022hand,saquib2019interactive,fribourg2021mirror}, we found that this was indeed possible if we allowed the presenter to `wear' the visualizations, or attach them directly to the body like costumes from theater. The visualization could influence the presenter by restricting certain body movements (similar to how different costumes could affect the actors movements), while the presenter simultaneously influence the visualization through their movements (costumes move as the actor moves).}

\subsection{\edits{A Tool to Implement \probes}}
\edits{To offer the viewer with \probes that place the presenter and the visualization on an equal footing, we needed to consider both their spatial layout, form, and interactions. Because there were no off-the-shelf tools that would allow us to create such \probes, we developed a new tool called the Body Object Model (BOM) to help create our probes. \textbf{While the tool itself is a means to an end, and not our contribution, we discuss it because it plays a key role in the process of creating our probes, \probes.} More detailed implementation details are in the Appendix.}

\begin{figure*}[h]
    \centering
    \includegraphics[width=\textwidth]{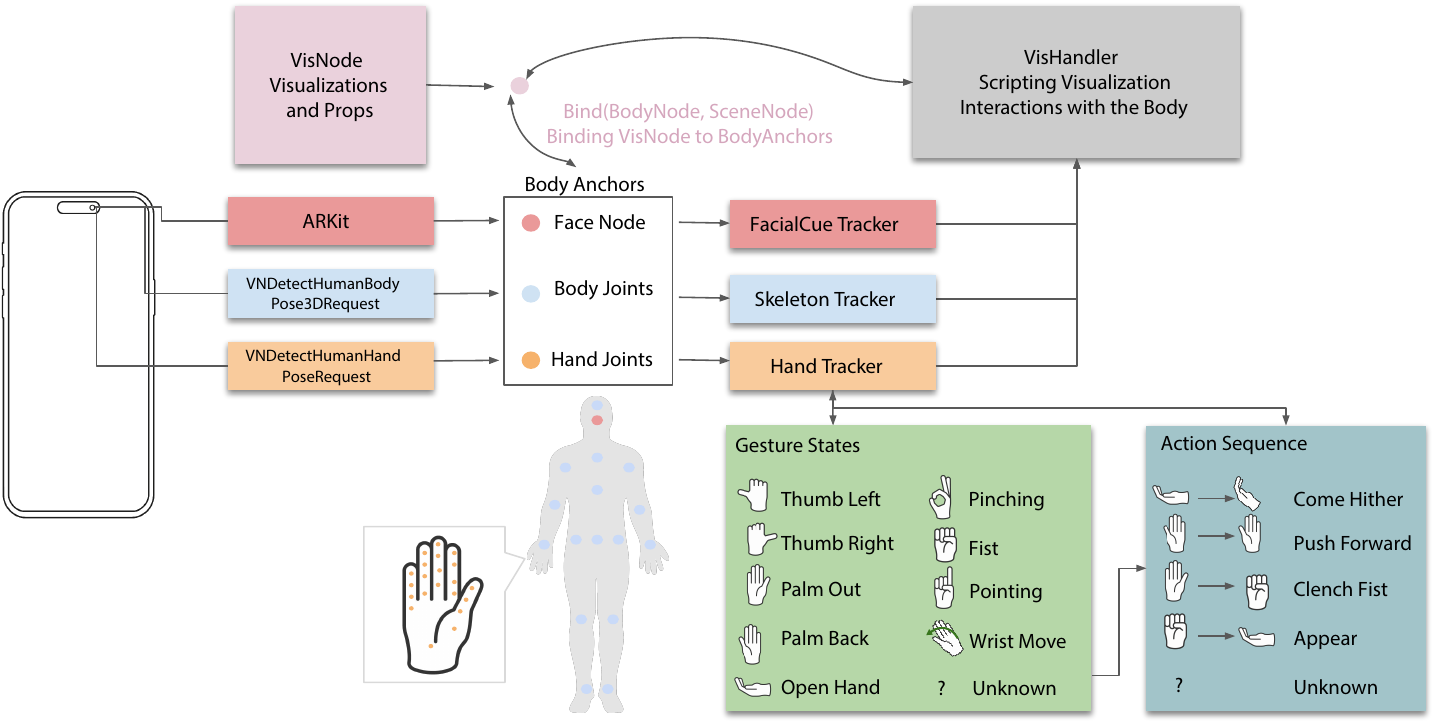}
    \Description[Body Object Model]{ Three arrows stemming from the self–facing camera to show where the data for generating Face Node, Body Joints, and Hand Joints are from. One of these arrows show how ARKit, coded in red, generates a red Face Node; the other how VNDetectHumanBodyPose3DREquest generates light blue body joints; and light orange for VNDetectHumanHandPoseRequest generating hand Joints. These face node, body joints, hand joints are encapsulated in a box called `body anchors’. There is a purple `VisNode’ box which generates a purple circle of either `Scatter Plot, Bar Chart, or Map Nodes’. There is a purple double arrow that connects the `body anchors’ with the `visNode’ with the caption `Bind(BodyNode, SceneBode), Binding VisNode to BodyAnchors’ to show that you can bind visNode with body anchors. Under the box demonstrates where the nodes are anchored to the body and the hands. The faceNode is on the center of the face, the blue nodes are at the top of the head, three points across the left, center, and right of the shoulder, the elbows, the spines, the wrists, the left, center, and right hip, the knees, and the ankles. The hand joints have one on the wrist and the rest have four for each joint of the fingers. There is an arrow from `Body Joints’ to `Skeleton Tracker’ to demonstrate that skeleton tracker manages those joints. There’s another from `hand joints’ to `Hand Tracker’ for a similar reason. There’s a double arrow stemming from `Hand Tracker’ to produce a green box of `Gesture States’. This green box is essentially a table of 9 gesture icons and their associated descriptions of what those hand gestures mean. There is `Thumb Left’, `Thumb Right’, `Palm Out’, `PalmBack’, `OpenHand’, `Pinching’, `Fist’, `Pointing’, and `Unknown’. There’s then another double arrow from `Gesture States’ to a grayish blue box called `Action Sequence’ to demonstrate that action sequences are created using `Gesture States’. There are 5 different action sequences and each of them show how the action is made using each of the gesture states. Hence gesture icon, arrow, gesture icon. The first one is a `Come Hither’ gesture which demonstrates a beckoning motion. `Push Forward’ is a series of `PalmOut’. 'Clench Fist’, is the user having their 'Palm Out’ and then making a `Fist’. The `Appear’ has the user make a `Fist’, and then `Open Hand’. The last one is an unknown state represented by a question mark. There is an arrow connecting Gesture States to Action Sequence. Then there is the gray `VisHandler’ box, with description ‘Scripting Visualisation Interactions with the Body’. Each of ‘FacialCue Tracker’, ‘Skeleton Tracker’ and ‘Hand Tracker’ has an arrow connecting to ‘VisHandler’. The `VisHandler’ also has a double arrow line that is purple with `VisNode’ to demonstrate that you can use it to bind gestures with the vis node.
}
    \caption{Demonstrates system implementation of Body Object Model (BOM). There are three main components: (1) \textsc{BodyAnchors} that represent locations in the presenter's body (2) \textsc{VisNode} which contain the visualizations and (3) \textsc{VisHandler} which is used to define relationships between the presenter and visualization using the \textsc{BodyAnchors} and \textsc{VisNodes}.}
    \label{fig:systemDiagram}
\end{figure*}

\edits{We implemented the Body Object Model using ARKit, an augmented reality package, on the iPhone as this enabled us to merge the virtual visualizations with the physical space of the presenter. To create a presentation where we could control the presenter-visualization relationship, we implemented three components: the \textsc{VisNode}, \textsc{BodyAnchors}, and the \textsc{VisHandler} (\autoref{fig:systemDiagram}). \textsc{VisNode}s are containers for 3D visualizations. The \textsc{BodyAnchors} represent locations on the presenter's body. The tracking system provides the location of the \textsc{VisNode} and \textsc{BodyAnchors} in 3D space. The \textsc{VisHandler} allows us to script the interactions between the visualization (\textsc{VisNode}) and the presenter (\textsc{BodyAnchor}).}

These building blocks enable the \edits{design of presentations with equitable relationships between the visualization and presentation with the} following three features: 
\begin{enumerate}
    \item[\textbf{F1.}] \textbf{Use of 3D Physical Space $\rightarrow$ C1}
    ARKit and the segmentation of the presenter from the physical background allows the designer to integrate 3D visualizations and place visualizations in different 3D locations. These new setups enhance the sense of depth and open new creative ways to integrate the physical surroundings with the presenter. The added z-dimension provides more space flexibility than 2D spaces, enabling more stagings in which the presenter's body does not compete for space with the visualization.  
    \item[\textbf{F2.}] \textbf{Body-vis Attachments $\rightarrow$ C2}  The performance designer can attach a \textsc{VisNode} to a \textsc{BodyAnchor}, almost like a wearable costume, and use the presenter's body as an integral part of the performance. These body-vis attachments enable simultaneously bidirectional interactions.  
    \item[\textbf{F3.}] \textbf{Body-vis Interactions $\rightarrow$ C3} The presenter can attach actions to the \textsc{VisNodes} and use their body to control or interact with them. 
\end{enumerate}

\subsection{\edits{Creating Four \probes}}
\edits{With the tools to create \probes in place, we then explored how different combinations of form and interactions (F1--F3) would affect the viewer's experience. We design four case \probes with varying conditions (C1--C3). To understand how the merged spaces of the visualizations and presenters affected the visualizations, for each \probe, we created its baseline equivalent representing a 2D videoconferencing style presentation. To investigate how the three-dimensionality of the visualizations would affect the viewing experience, we made two \probes with visualizations that looked two-dimensional (\injuryVis, \walmartVis) and two \probes with three-dimensional visualizations (\airplaneVis, \napoleonVis). To understand how different types of interactions may affect the viewer, we strategically created one \probe where the visualization induced the presenter to move around it (\airplaneVis), one where it was evident the presenter moved the visualization (\napoleonVis), and one where the presenter and visualization were bound to each other physically and interaction wise (\injuryVis).}

\edits{The content for each \probe was selected with these conditions in mind. We narrowed our search scope to infographics ~\cite{injuriesExampleVis,minardExampleVis,planeExampleVis}, because they are known for their simple and engaging data narratives~\cite{bateman2010useful, harrison2015infographic, li2014chart}. This eliminated the need to craft a captivating storyline for the viewers from scratch. To refine the design of our \probes and ensure that they were engaging from a viewer's perspective, we conducted internal critiques with 15 HCI researchers over 4 months with alternative designs~\cite{tohidi2006getting} before evaluating them with viewers in public.}

The following paragraphs provide a brief description of each InfoVid, the reason behind its selection, and what elements reflect the conditions C1--C3. Additionally, we describe how we make the baseline, or each InfoVid's equivalent in a videoconferencing format. A summary on how the conditions C1--C3 are applied is available in \autoref{tab:visDesignControls}.
 
\begin{table*}[h]
\begin{tabular}{@{}lccc@{}}
\toprule
\begin{tabular}[c]{@{}c@{}}Visualization \\ Name\end{tabular} & \begin{tabular}[c]{@{}c@{}}C1. Visualizations Use \\ 3D Physical Space\end{tabular} & \begin{tabular}[c]{@{}c@{}}C2. Body-Vis Attachments for \\ Simultaneous, Bidirectional Interactions\end{tabular} & \begin{tabular}[c]{@{}c@{}}C3. Unilateral Body-Vis \\ Interactions by the Presenter\end{tabular} \\ \midrule
\staticVis                                                     & -                                                                                                            & -                                                                                                        &                                                                                                          \\
\airplaneVis                                                      &\checkmark                                                                                                            & -                                                                                                        & -                                                                                                        \\
\napoleonVis                                                   & \checkmark                                                                                                            & -                                                                                                        & \checkmark                                                                                                        \\ 
\injuryVis                                                     & -                                                                                                            & \checkmark                                                                                                        & \checkmark                                                                                                        \\
\walmartVis*                                                    & -                                                                                                            & \checkmark                                                                                                        & \checkmark                                                                                                        \\
\bottomrule
\end{tabular}

\caption{Describes how \textsc{\probes} apply the three interactions for each visualization design.  Comparing each visualization with a baseline version of itself helps investigate the effects of merged spatial layout. *\walmartVis is intentionally designed to have no major benefits from the body binding to provide a sample case that merely does a body-vis binding.}
 \Description[Condition Table]{Shows a table with four columns are five rows, each row with a visualization name. The Under the first column, ‘Physical Surroundings Contextualizes Visualization (Criteria 1), Baseline is unchecked, AirplaneVis is checked, WalmartVis is unchecked, InjuryVis is unchecked, and NapoleanVis is checked. Under ‘Performer Body as Contextualizes Visualization (Criteria 2), Baseline is unchecked, AirplaneVis is unchecked, WalmartVis is unchecked, InjuryVis is checked, and NapoleanVis is unchecked. Under ‘Performer Body-Vis Binding Enhances Narrative (Criteria 3), Baseline is unchecked, AirplaneVis is unchecked, WalmartVis is checked with with one asterisk, InjuryVis is checked, and NapoleanVis is checked with double asterisk.}
    \label{tab:visDesignControls}
\end{table*}

\paragraph{\textbf{\airplaneVis (C1)}} Based on a commercial airplane crash infographic~\cite{planeExampleVis}, this uses a 3D airplane and the physical surroundings (C1), such as the sky, to accentuate the sense of depth and present a scenario where the presenter must step to the side of the visualization because it takes 3D space (\autoref{fig:teaser}). 

\begin{figure*}[h]
    \centering
    \includegraphics[width=0.7\textwidth]{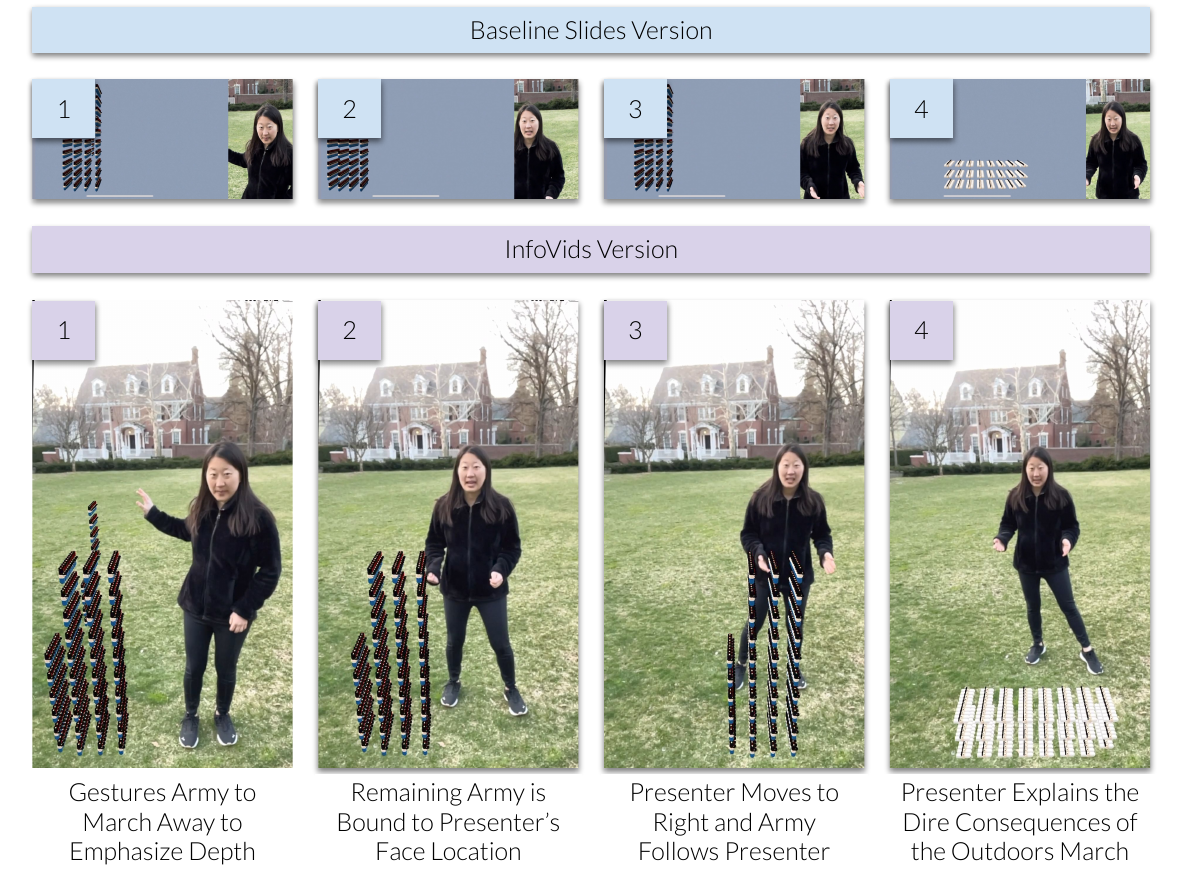}
    \Description[NapoleonVis]{Shows two sections, one section has a blue banner named ‘Baseline Slides Version’, beneath it four horizontal slideshow images labeled with blue labels that mark 1, 2, 3 and 4. Each blue label image shows on the left the NapoleonVis, a tall wall of orange, virtual dots that is animatable representing the size of french soldiers, and on the right a vertical video space for the presenter. The background of the NapoleonVis is gray. From image 1 through 4, the presenter gestures away from the screen, moves towards the left, then right, and eventually back to the center. The animation shows a portion of the army moving back, then the entire army moving left and right, and lastly the army laying horizontal on the ground, colored in white to represent casualties. The second section has a purple banner named ‘InfoVids Version’, beneath it are four vertical images labeled with purple labels that mark 1, 2, 3 and 4. The presenter moves around from images 1 through 4 and is standing on a field outside. In image 1, the presenter stands on the right hand side of NapoleonVis and gestures away from the NapoleonVis. A portion of the army moves further away from the camera. The label reads ‘Gestures Army to March Away to Emphasize Depth’. In image 2, the presenter shifts towards the left while holding her first, the label reads ‘Remaining Army is Bound to presenter’s Face Location’. In image 3, the presenter shifts towards the right, and the NapoleonVis follows her body. The label reads, ‘presenter Moves to Right and Army Follows presenter’. In image 4, the presenter shifts back to the center, the armies are rotated by 90 degrees, now laying flat on the ground, colored in white to represent casualties. The label reads, ‘presenter Explains the Dire Consequences of the Outdoors March’.
}
    \caption{\napoleonVis shows the consequences of Napoleon's march. The two compared versions are shown: one where slides with the animation is positioned to the left of their presentation (top), and one where the animation is positioned in the presenter's environment with \probes (bottom). In frames 2 and 3, the presenter binds the French army to their body to control army movements directly. This presents an engaging way to perform the chase of the French after the Russian army (\autoref{tab:visDesignControls} Criteria 3). The background reinforces the `outdoors' setting of the march and provides depth when the initial army parts into the distance in Frame 1 (Criteria 1). These design choices made \textsc{\napoleonVis} significantly more enjoyable to watch for the viewer.}
    \label{fig:napoleon_vis}
\end{figure*}

\begin{figure*}[p]
    \centering
    \includegraphics[width=0.7\textwidth]{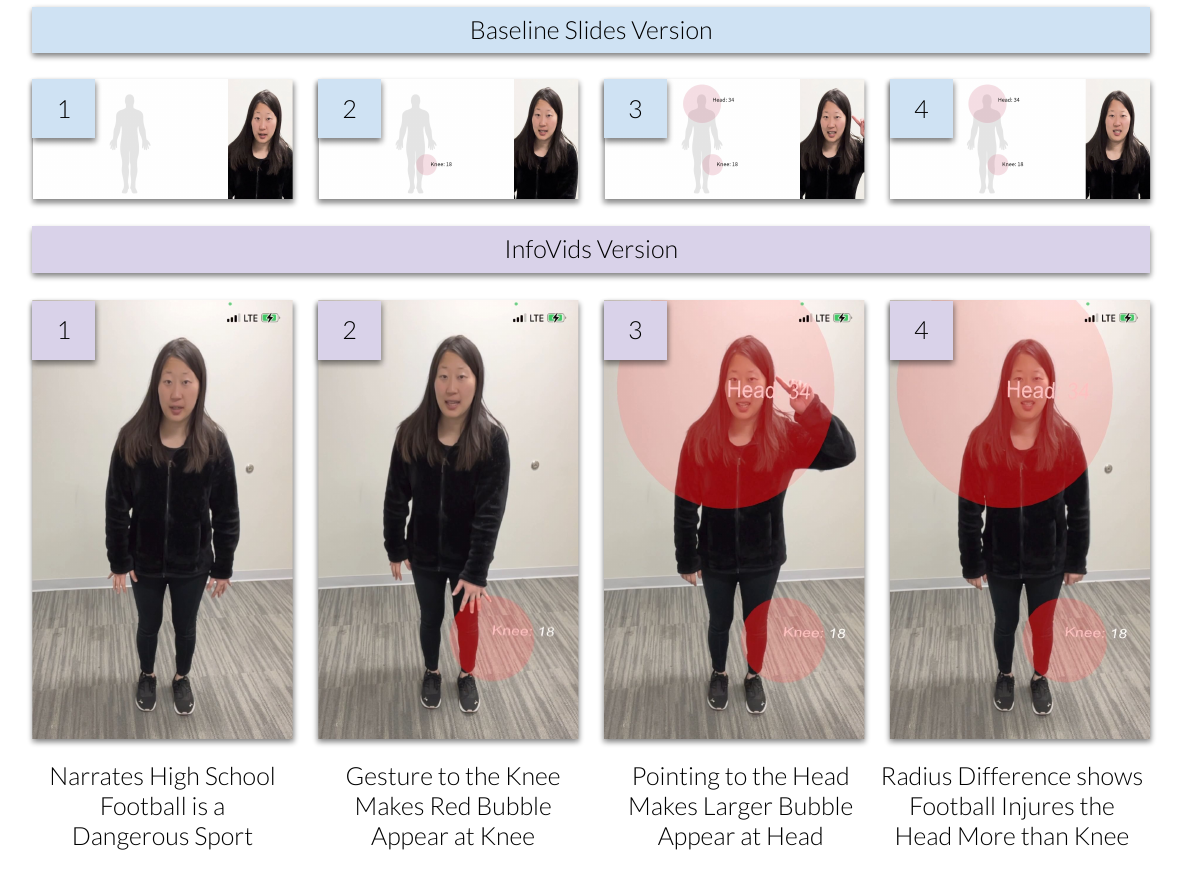}
    \Description{ Shows two sections, one section has a blue banner named ‘Baseline Slides Version’, beneath it four horizontal slideshow images labeled with blue labels that mark 1, 2, 3 and 4. Each blue label image shows a gray diagram of a person against white background on the left, and a vertical video space for the presenter on the right. The presenter moves around slightly from images 1 through 4. In label 1, the figure is unlabelled. Starting from label 2, a red, half-transparent red circle appears on the knee of the figure. It is labeled ‘Knee: 18’. In image 3, a larger red circle appears around the head of the figure, labeled ‘Head: 34’. The presenter is posting towards their head with their left index finger. In the final image, the presenter puts their arm down and the visualization stays the same, indicating the end of the performance. The second section has a purple banner named ‘InfoVids Version’, beneath it are four vertical images labeled with purple labels that mark 1, 2, 3 and 4. In each image, the presenter is shown full body standing in a room. In image 1, the presenter appears to be explaining. The caption reads, ‘Narrates High School Football is a Dangerous Sport’. In image 2, the presenter gestures to the left knee with their left hand, and a half–transparent red circle appears at their knee, with label ‘Knee: 18’. The circle is about the size of the presenter’s head. The caption reads ‘Gesture to the Knee Makes Red bubble Appear at Knee’. In image 3, the presenter uses their left index finger to point at their head. Another red circle appears around the head. It is extremely big, about 5 times the size of the presenter’s head. The caption reads ‘Pointing to the Head Makes Larger Bubble Appear at Head’. In image 4, the presenter puts their left arm down. The caption reads, ‘Radius Difference shows Football Injures the Head More than Knee’.}
    \vspace{-0.25cm}
    \caption{\injuryVis uses the presenter's body as context as the presenter points to their \textit{own} body to explain football injury statistics (bottom). These body-vis bindings engage the viewers by presenting a more personalized way to display data (\autoref{tab:visDesignControls} C3 \& C4). Viewers express mixed feelings with \injuryVis's ability to be enjoyed more. This is partially attributed to mismatches in participant expectations and the capability of a regular slides (top) presentation to present information in a more familiar way.}
    \label{fig:injury_vis}
\end{figure*}

\begin{figure*}[p]
    \centering
    \includegraphics[width=0.7\textwidth]{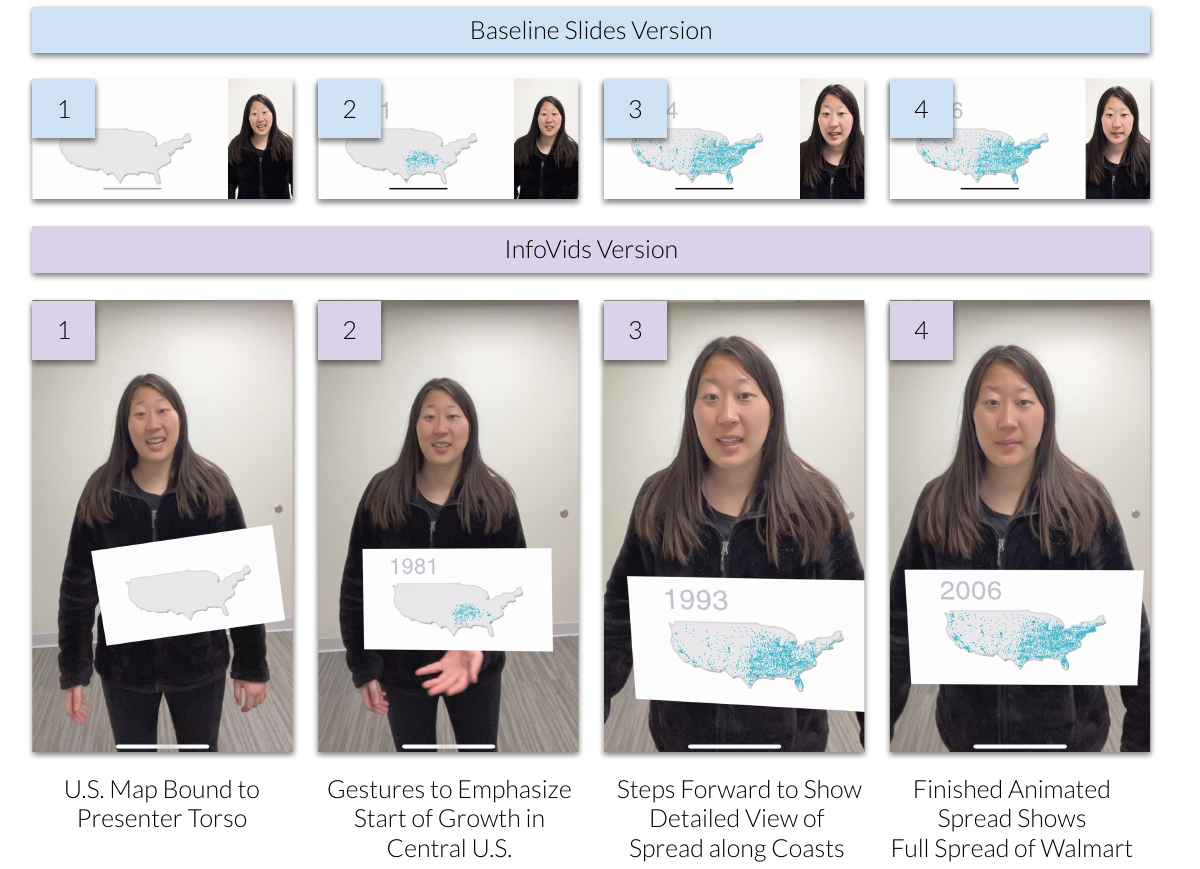}
    \Description[]{Shows two sections, one section has a blue banner named ‘Baseline Slides Version’, beneath it four horizontal slideshow images labeled with blue labels that mark 1, 2, 3 and 4. Each blue label image shows a rectangular US map with white background on the left, and a vertical video space for the presenter on the right. The presenter gets closer to the camera from images 1 through 4. In label 1, the US map is empty. Starting from label 2, blue dots showing Walmart locations in the US begin to appear. Between 2 and 4, the number of blue stores increases as they spread out from the South towards the West Coast. The second section has a purple banner named ‘InfoVids Version’, beneath it are four vertical images labeled with purple labels that mark 1, 2, 3 and 4. In each image, the presenter is shown from knee up. A rectangular US map with white background is attached to the torso of the presenter. In image 1, the caption reads ‘U.S. Map Bound to presenter Torso’. In image 2, the presenter is using their left hand to trigger the animation, which triggers the date 1981 to appear on the map as well as little blue dots representing store locations. The caption reads ‘Gestures to Emphasize Start of Growth in Central U.S’. In image 3, the caption reads ‘Steps Forward to Show Detailed View of Spread along Coasts, representing the image content, with the map dated 1993. In Image 4, the date is 2006 on the map and the caption reads ‘Finished Animated Spread Shows Full Spread of Walmart’.}\vspace{-0.25cm}
    \caption{\textsc{\walmartVis}' binding of the visualization to the body serves no major benefits to the narrative (\autoref{tab:visDesignControls}). It frees the hands of the presenter and lets them walk forward to view the visualization closer within the capturing camera. While viewers overall believe presenter immersion, engagement, and sense of co-location with the presence is overall higher than in the baseline, body movement constantly tilts the map, making the performance less enjoyable to watch.}
    \label{fig:walmart_vis}
    \vspace{-0.5cm}
\end{figure*}

\paragraph{\textbf{\minardVis (C1, C3)}} This references Charles Minard's depiction of the catastrophic march across Russia in 1812 and Numberphile's presentation~\cite{kosara2013storytelling, minardNumberphile, minardExampleVis}. The 3D army and the physical surroundings, the grass, reinforces the outdoor setting of the event as the army parts into the z-axis (\autoref{fig:napoleon_vis}, Frame 1). Mid performance (\autoref{fig:napoleon_vis}, Frames 2 \& 3), the presenter uses their full-body to move back and forth the French army to simulate the French following the Russians (C3).

\paragraph{\textbf{\injuryVis (C2, C3)}} A direct homage to the New York Times infographic~\cite{injuriesExampleVis}, \injuryVis fuses the presenter and the visualization with multiple body-vis bindings, making the presenter inseprable from the visualization, and act as an integral narrative device of the visualization (C2). The presenter points to their own body (C3) to demonstrate the dangers of football. Translucent red bubbles appear on the presenter's ankle and head to indicate the number of injuries at that region of the body (\autoref{fig:injury_vis}).

\paragraph{\textbf{\walmartVis (C2, C3)}} The animated geoplot, a replica of Bostock's D3.js animation~\cite{walmartExampleVis}, is a controlled counterexample. It is crafted to understand when body-vis bindings are useful or harmful for viewers. The animated map is bound to the presenter's torso, but serves little narrative purpose, unlike the \injuryVis.

\begin{figure*}[b]
    \centering
    \includegraphics[width=0.90\textwidth]{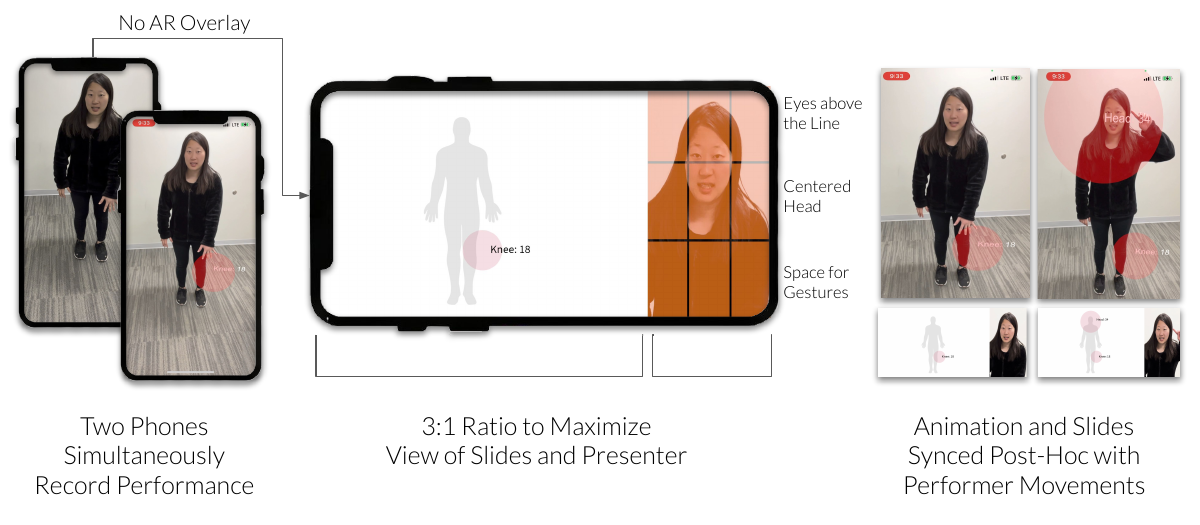}
    \Description[Making the Baseline Video Format]{Shows the three sections representing steps of creating Baseline Video Format. First section is labeled ‘Two Phones Simultaneously Record Performance’. Two iPhone interfaces are shown, each with a presenter standing in a room. One screen has an AR overlay and the other does not. The one without AR overlay is connected to the second section via an arrow, with the label, ‘No AR Overlay’. The second section is labeled ‘3:1 Ratio to Maximize View of Slides and presenter’. It shows a horizontal iPhone screen. The leftmost 3/4 of the screen is occupied by a white slide with a gray figure and a semi-transparent red circle around the knee. The other 1/4 is occupied by the video taken from the no AR overlay iPhone video from section 1. On this video there is a semi-transparent orange overlay and black grid lines that separates the video into 9 sections. The three rows of grids are labeled ‘Eyes above the Line’, ‘Centered Head’ and ‘Space for Features’. The presenter’s eyes are lined up within the first row of grids, etc. The third section is labeled ‘Animation and Slides Synced Post-Hoc with presenter Movements’. It shows four images, the top two are screenshots from the video with AR Overlay, taken at different time frames. The bottom two images are screenshots from the synced up slideshow performance.}%
    \vspace{-0.5cm}  \caption{\textbf{Making the Baseline Video Format} To use the same rendition, or takes, of the performance, two phones are used simultaneously to record the presenter---one uses \textsc{\probes} and the other is absent of the augmented visualizations. The latter is used to sync the presenter with the slides and animation. presenter body language and visualization views are maximized with a 3:1 slide format and use of rule-of-thirds, a video composition technique to direct viewer attention to the presenter.}
    \label{fig:editing_process}
\end{figure*}

\paragraph{\textbf{Baseline}}
The baseline is the equivalent InfoVid as a 2D videoconferencing style presentation with slides. Comparing the baseline with the InfoVids will allow us to understand the effects of the new stagings enabled from the added flexibility of 3D space (F1). We minimized confounding factors between the baseline and \probes by using the same take of the performance and scripting a common body language that worked for both formats. Given the limited screen size of a mobile phone, we chose a horizontal video orientation for the baseline such that the presenter's upper body, including their face and hands, would remain visible. To make aesthetically pleasing layouts, we used a video composition technique called the rule-of-thirds. We summarize this in \autoref{fig:editing_process}. However, making a baseline format that enables a fair comparison between the baseline and the InfoVids using the same performance takes is challenging and requires a detailed discussion of its own. Therefore, we provide a detailed discussion of the trade-offs and the reasons behind our choices in the Appendix.

\section{A Comparative Probing with \probes}
We conducted street interviews to evaluate the \probes with members of the public. To poll diverse people, we followed Denning et al.'s methodology---interviews were conducted at three different cafes and a mall because of their \squote{reasonable throughput of traffic} and ability to \squote{attract different demographics}~\cite{denning2014situ}. Two interviewers worked in parallel each for 10 hours to ask potential participants whether they would like to participate in a 15-minute evaluation of visual performances in exchange for a pastry. Demographic information was not collected to encourage interviewees to respond more and to provide honest feedback. Out of respect of people's time and space, we only approached people who appeared to be waiting or not preoccupied. A total of 74 people were approached, of which 44 declined to participate. Subsequent analyses are based on the 30 participants who voluntarily consented to participate.

\subsection{Interview Procedure}
Participants were shown 4 pairs of performances on a smartphone. Each pair showed the same topic with one in a video conferencing format and another with \probe. As shorthand, we called the baseline condition the `slides' version and the latter the `non-slides' one. Both videos had $886\times1920$ resolution and a maximum length of 35 seconds. To minimize ordering and recency bias, we took random permutations of the 4 pairs of videos and randomly chose which version of the presentation was shown first.

After each pair of viewings, participants filled an anonymous online 9-question survey on a laptop. Question orderings were randomized to mitigate ordering bias. We ask the following 9 questions in the survey:
\begin{enumerate}
    \item \textit{Which presentation makes the presenter look more immersed with the visualization?}
    \item \textit{Which presentation makes the presenter look more engaged with the visualizations?}
    \item \textit{Which presentation makes it more believable is that the presenter is in the same room as the visualizations?}
    \item \textit{Which presentation does the presenter look more natural, body movement wise?}
    \item \textit{Which presentation is more enjoyable to watch?}
    \item \textit{Which presentation style is better for telling a story?}
    \item \textit{Which presentation style is better for understanding the information?}
    \item \textit{For the NON-SLIDES version, did you view the presenter or the visualization more? } 
    \item \textit{For the SLIDES version, did you view the presenter or the visualization more?}
\end{enumerate}

\edits{The questions are designed as a two-alternative forced choice test (2AFC), but participants were presented with a 6-point Likert scale to discourage random selection bias. The analysis, however, treats the answers as a 2AFC, as this was the intended framework from the onset of the study design.} After all 4 surveys, we conducted a short semi-structured interview to ask what elements in the performances affected their decisions. The interviewer transcribed notes in real-time to record the participant responses.

\subsection{Method of Analyses}
There were four major parts to the mixed methods analyses. First, to analyze the overall efficacy of \probes for each visualization, we first investigated participants' overall preferences between the baseline and the \probe and the presenter and visualization. For each of the 9 survey questions for a given visualization, a binomial test was conducted with $(\alpha=0.05)$, because each question was designed as a 2AFC test with 2 possible outcomes. The null hypothesis $H_0$ assumed both outcomes were equally likely $(p=0.5)$. Rejecting the null hypothesis signified the observed preference over one outcome or the other was not random and was statistically significant. 

Next, to understand if the different survey results among visualization types were significant, we ran the Friedman's Chi-square tests for each of the question. The Friedman test ($\alpha=0.05$) was used because the same viewers ordinally rated each question using a 6-point Likert scale on 4 different conditions. The $H_0$ in this case assumes there is no difference in the medians, or that the results across different visualization types were comparable. To further investigate whether a pair of groups significantly different from each other, we ran the Wilcoxon Signed-Rank test to see if the distribution of differences between the two groups were approximately normal, else we ran a weaker signed test.  

We also analyzed how the new staging format introduced by \probes affects viewer attention to the presenter and visualization prop. For each visualization type, we twice asked participants whether they viewed the presenter or the visualization more---once for the \probe and another for the baseline. For each visualization, we compare whether the \probe made significant differences in visualization-presenter attention from the baseline. As we have fewer than 50 participants and paired, binary outcomes (visualization/presenter), we conduct the Fisher's exact test with $\alpha=0.05$ and summarize the results in \autoref{fig:breakdown_vis_or_presenter}. To evaluate how many viewers switched their focus more to the presenter with \probes, in comparison to the baseline, we also analyze the directionality of change using contingency tables.  

Lastly, to understand the factors that affected the viewer's decisions, we thematically analyzed the semi-structured interviews using deductive coding~\cite{braun2006using}. Given the short interviews, only three iterations were needed to generate the following 9 themes: captivating, aesthetics, distracting, attention split, helpful for information understanding, breaks in presence, body language and interaction, body as context, and situational context.

\begin{figure*}[p]
\includegraphics[width=0.8\linewidth,height=19cm]  {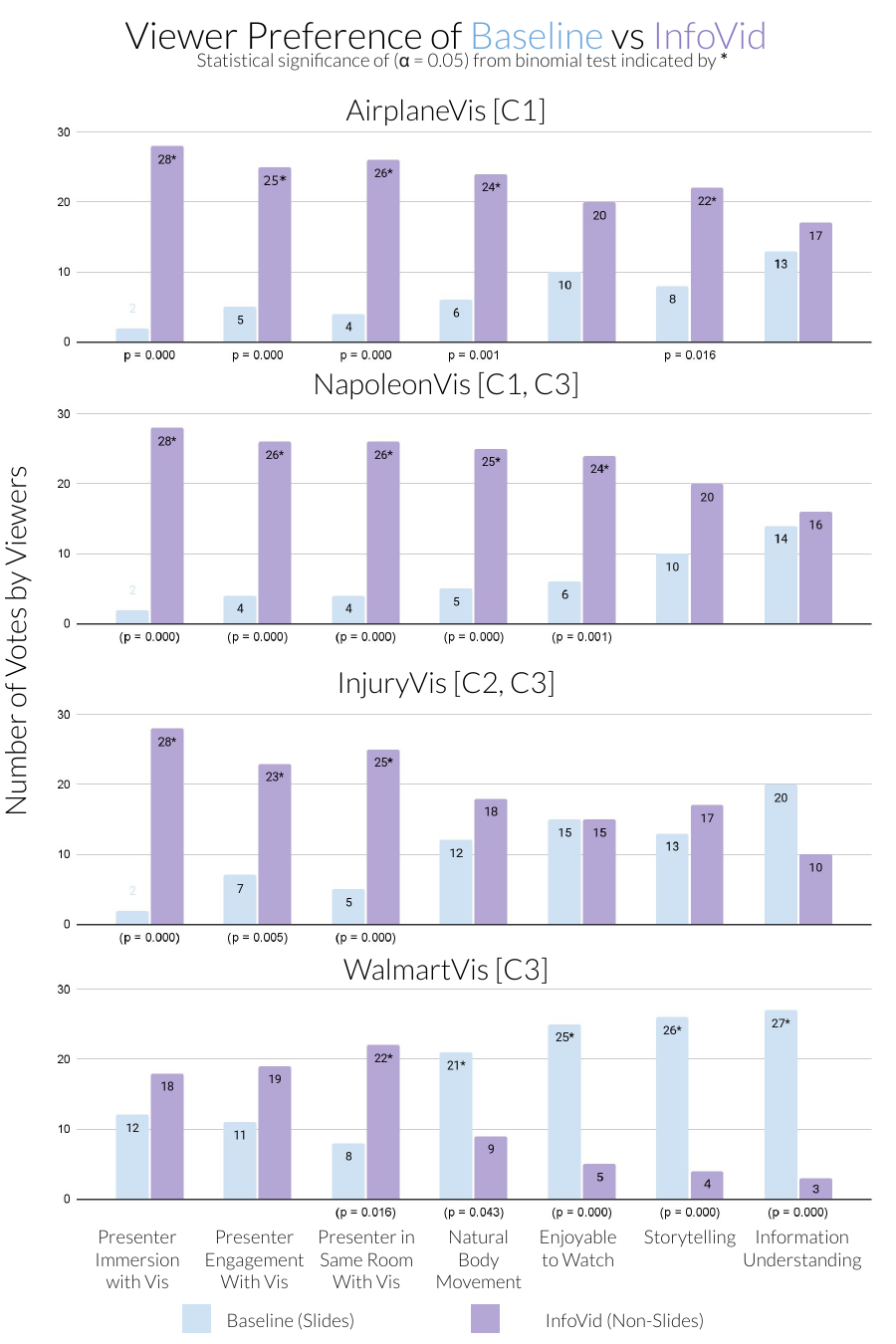}
  \Description[]{: Shows four charts of view preference of Baseline vs InfoVids between the four sets of videos, AirplaneVis, NapoleonVis, InjuryVis and WalmartVis. Underneath the Heading, ‘Viewer Preference of Baseline vs InfoVids’, it reads ‘Statistical significance of (a = 0.05) from binomial test indicated by *’ For each chart, the y axis is ‘Number of Votes by Viewers, going from 0 to 30, and the x axis are categories questions asked, including ‘Presenter Immersion with Vis’, ‘Presenter Engagement with Vis’, ‘Presenter in Same Room with Vis’, ‘Natural Body Movement’, ‘Enjoyable to Watch’, ‘Storytelling’ and ‘Information Understanding’. For each chart, Baseline(Slides) votes, in the form of a bar, are represented by sky blue and the first chart. InfoVids(Non-Slides) votes are represented by purple bars. In the first chart, ‘AirplaneVis’, presenter Immersion with Vis scores a Baseline of 1 and InfoVids of 28, with a p value of 0.000. Presenter Engagement with Vis scores a Baseline of 5 and InfoVids of 25, with a p value of 0.000. Presenter in Same Room with Vis scores a Baseline of 4 and InfoVids of 26, with a p value of 0.000. Natural Body Movement scores a Baseline of 6 and InfoVids of 24, with a p value of 0.001. Enjoyable to Watch scores a Baseline of 10 and InfoVids of 20. Storytelling scores a Baseline of 8 and InfoVids of 22, with a p value of 0.016. Information Understanding scores a Baseline of 13 and InfoVids of 17. In the second chart, ‘NapoleonVis’, presenter Immersion with Vis scores a Baseline of 1 and InfoVids of 28, with a p value of 0.000. Presenter Engagement with Vis scores a Baseline of 4 and InfoVids of 26, with a p value of 0.000. Presenter in Same Room with Vis scores a Baseline of 4 and InfoVids of 26, with a p value of 0.000. Natural Body Movement scores a Baseline of 5 and InfoVids of 25, with a p value of 0.000. Enjoyable to Watch scores a Baseline of 6 and InfoVids of 24, with a p value of 0.001. Storytelling scores a Baseline of 10 and InfoVids of 20. Information Understanding scores a Baseline of 14 and InfoVids of 16. In the third chart, ‘InjuryVis’, presenter Immersion with Vis scores a Baseline of 1 and InfoVids of 28, with a p value of 0.000. Presenter Engagement with Vis scores a Baseline of 7 and InfoVids of 23, with a p value of 0.005. Presenter in Same Room with Vis scores a Baseline of 5 and InfoVids of 25, with a p value of 0.000. Natural Body Movement scores a Baseline of 12 and InfoVids of 18. Enjoyable to Watch scores a Baseline of 15 and InfoVids of 15. Storytelling scores a Baseline of 13 and InfoVids of 17. Information Understanding scores a Baseline of 20 and InfoVids of 10. In the fourth chart, ‘WalmartVis’, presenter Immersion with Vis scores a Baseline of 12 and InfoVids of 18. Presenter Engagement with Vis scores a Baseline of 11 and InfoVids of 19. Presenter in Same Room with Vis scores a Baseline of 8 and InfoVids of 22, with a p value of 0.016. Natural Body Movement scores a Baseline of 21 and InfoVids of 9, with a p value of 0.043. Enjoyable to Watch scores a Baseline of 25 and InfoVids of 5, with a p value of 0.000. Storytelling scores a Baseline of 26 and InfoVids of 4, with a p value of 0.000. Information Understanding scores a Baseline of 27 and InfoVids of 3, with a p value of 0.000.}
  \caption{\textbf{Participants were asked about which version of presentation format immersed and engaged the presenter with the visualization more, felt like the presenter was in the same room as the visualization, had more natural body movement, was more enjoyable to watch, better storytelling, and better information understanding. Across all four visualizations, participants chose the \probe version for immersion, engagement, and being in the same room as the visualization. Their responses to the other four questions depended on the visualization, as shown in the charts.}}
\label{fig:breakdown_slides_or_nonslides}
\end{figure*}

\section{Findings and Lessons Learned from the Viewer's Perspective}
\edits{By asking participants to compare among different \probes and between the baseline and the \probes, we aim to understand the following questions: (1) How do \probes affect the viewing experience compared to traditional videoconferencing formats? (2) Does the merging of spaces make viewers focus on the presenter more than the visualization (survey Q8-9)? (3) How do the different conditions (C1--C3) in the \probes affect their viewing experiences (survey Q1-Q7)? We hope the answers to these questions can better guide the future design and presentation of \probes-like visualizations.} 

\begin{figure*}[h]
    \includegraphics[width=\textwidth]{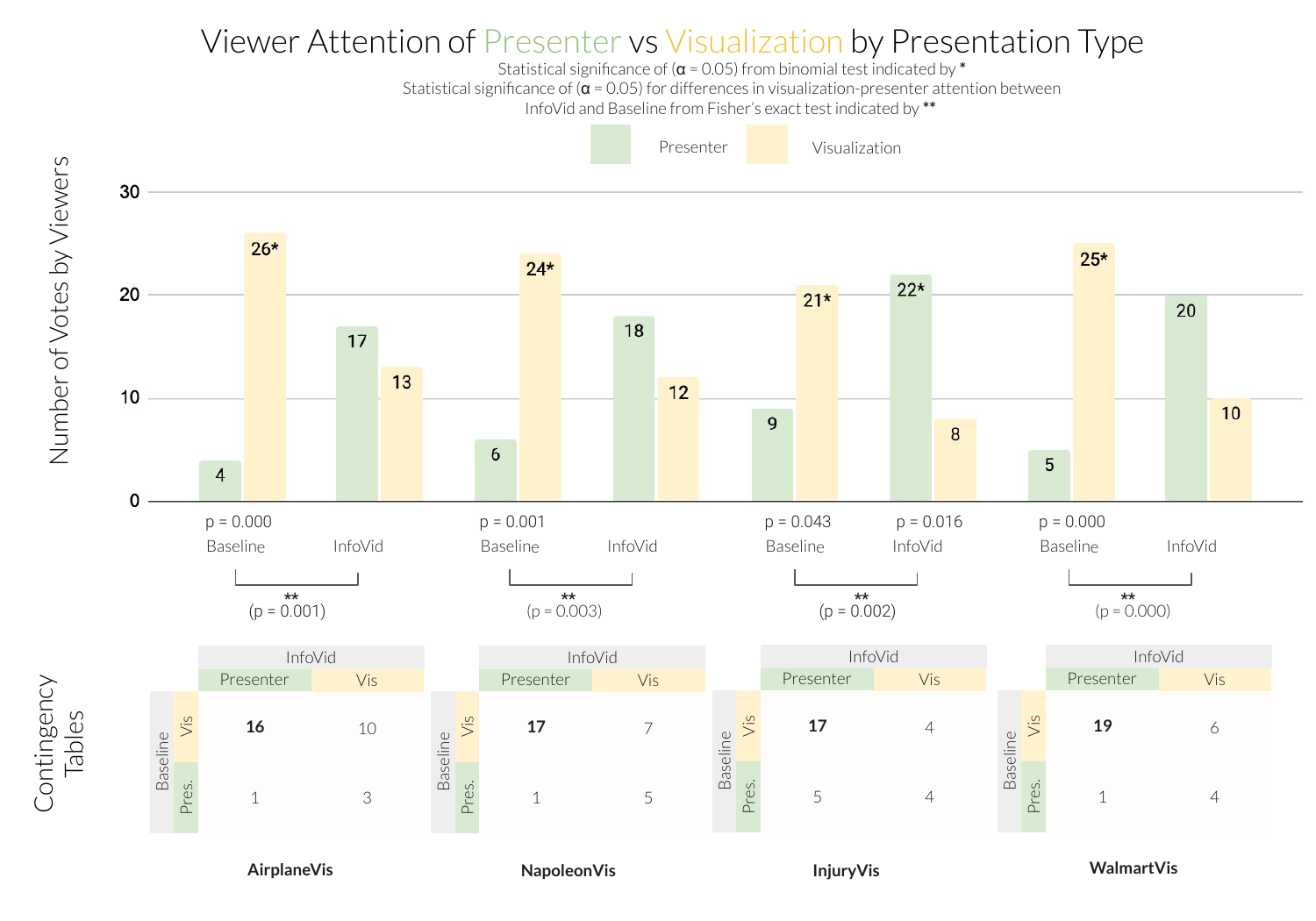}
    \Description{  Shows four bar charts and four contingency tables for the four sets of visualizations. Under the heading, ‘Viewer Attention of presenter vs Visualization by Presentation Type’, it reads ‘Statistical significance of (a = 0.05) from binomial test indicated by *’, “Statistical Significance of (a = 0.05) for differences in visualization-presenter attention between InfoVids and Baseline from Fisher’s exact test indicated by **’ presenter is indicated by green and visualization is indicated by yellow. For each bar chart, the y axis is Number of Votes by Viewers ranging from 0 to 30, the x axis categories are Baseline and InfoVids. For AirplaneVis, the Bar chart shows that Baseline scores a presenter score of 4, Visualization score of 26*, p value of 0.000. InfoVids scores a presenter score of 17 and Visualization score of 13. The p value for differences between Baseline and InfoVids is 0.001. For NapoleonVis, the Bar chart shows that Baseline scores a presenter score of 6, Visualization score of 25*, a p value of 0.001. InfoVids scores a presenter score of 18 and Visualization score of 12. The p value for differences between Baseline and InfoVids is 0.003. For InjuryVis, the Bar chart shows that Baseline scores a presenter score of 9, Visualization score of 21*, a p value of 0.043. InfoVids scores a presenter score of 22*, Visualization score of 8, p value of 0.016. The p value for differences between Baseline and InfoVids is 0.002. For WalmartVis, the Bar chart shows that Baseline scores a presenter score of 5, Visualization score of 25*, a p value of 0.000. InfoVids scores a presenter score of 20, Visualization score of 10. The p value for differences between Baseline and InfoVids is 0.000. Underneath the bar charts shows four contingency tables. Each table has a horizontal heading ‘InfoVids’ spanning across two columns. The first column is labeled ‘presenter’, the second column is labeled ’Vis’. The Vertical heading is ‘Baseline’, spanning across two rows. The first row is labeled ‘Vis’ and the second labeled ‘Pres.’, short for presenter. For AirplaneVis, the value at row 1 column 1 is 16, value at row 1 column 2 is 10. Value at row 2 column 1 is 1, and value at row 2 column 2 is 3. For NapoleanVis, the value at row 1 column 1 is 17, value at row 1 column 2 is 7. Value at row 2 column 1 is 1, and value at row 2 column 2 is 5. For InjuryVis, the value at row 1 column 1 is 17, value at row 1 column 2 is 4. Value at row 2 column 1 is 5, and value at row 2 column 2 is 4. For WalmartVis, the value at row 1 column 1 is 19, value at row 1 column 2 is 6. Value at row 2 column 1 is 1, and value at row 2 column 2 is 4.
}
    \vspace{-0.75cm}   \caption{\textsc{\probes} disrupt this traditional visual flow as demonstrated by overall ratio of presenter to visualization ratio for the \probes. Binomial tests indicate statistical significance for \textsc{\injuryVis} because the presenter and visualization are infused with each other by design. Bolded in the contingency tables are the number of converts, or viewers who focused on the visualizations in the baseline but the presenter with \probes.}
    \label{fig:breakdown_vis_or_presenter}
\end{figure*}

According to survey results, \probes enhances the stories and makes the experience more enjoyable for viewers in some cases than the baseline. From the rationale given, the viewers perceived the presenter to use more of their natural body movements and looked more immersed, engaged, and physically present with the visual props (\autoref{fig:breakdown_slides_or_nonslides}). While viewers primarily focus on the visualizations in the baseline, \probes make the viewer focus on the presenter more by blending the presenter into the same space with the visual props (\autoref{fig:breakdown_vis_or_presenter}). 

\paragraph{\edits{\textbf{Body Language is All You Need: \probes Enable Engaging and Enjoyable Full-Body Performances}}}

The results suggest that \probes were more engaging and enjoyable to viewers because the \edits{layout designed to foster an equal standing between the visualization and presenter} encouraged the use of the full-body with the blended space. While the baseline had the same animations and performance takes, synced to the presenter's movements, the presenter's body was cut off, \edits{making the presentation visualization-centric by design}. Participants \pquote{couldn't really see her [the presenter's] whole body}{P8}. The baseline format prevented viewers from fully seeing the presenter's actions directly related to the visualizations, such as pointing to different parts of the plane. 

However, \edits{by co-locating the presenter and visualization in the same 3D space (C1), we changed the relationship between them, and could send the message to the viewers that the presenter's body holds more meaning in the presentations than traditional ones.} The presenter's body, scripted to move to the side for a crashing plane in \airplaneVis, emphasizes that the plane is taking up the presenter's 3D space. The presenter's body is the visualization itself in \injuryVis and a symbol for the Russians that the French were following in \napoleonVis. As P14, a strong advocate for \probes, states: 

\mylongquote{Even for \walmartVis, body language is all you need; you don't need different languages---French, English, Spanish---it's understandable for anyone it's eye grabbing and it just pulls your attention and makes you focus on the presenter all the time and it's engaging.} 

The body-vis bindings in \minardVis(C3) and \injuryVis(C2) emphasized the use of the full-body by enhancing direct interactivity with the visualizations. As a result, participants \pquote{liked how the soldiers [from \minardVis] were going back and forth and [that] the presenter was moving with them}{P1}. Some felt it \pquote{connected the story with the movement}{P22} of the presenter. \injuryVis was engaging because the \pquote{presenter is pointing at their own body}{P25} and \pquote{refer[ring] to her own body}{P23} and some were even \pquote{wow[ed]}{P29}. As P17 states: \mylongquote{The body one, it was really appropriate---what better way to talk about the body than to use the body?}

Even \airplaneVis (C1) was preferred over the baseline because of \pquote{the way the presenter moved to the side and occupied the space of the frame...made it more cohesive}{P17}. Moving to the side of the frame was scripted, but did not involve any direct body-vis bindings or interactions (C2, C3). Thus, this indicates that the presenter's full-body on its own acts a critical narrative device and occluding the body in the baseline makes the presentations less engaging. 

We corroborate this sentiment in the quantitative findings (\autoref{fig:breakdown_slides_or_nonslides}). Viewers found presenters in \airplaneVis, \napoleonVis, and \injuryVis to look more natural and better at storytelling than the baseline.  Furthermore, the majority of viewers found the \airplaneVis and \napoleonVis \probes to be more enjoyable to watch than the baseline.

\paragraph{\textbf{\probes Reduce Split Viewer Attention and Increase Attention on the Presenter}}
Viewers preferred \probes because the blending of props and presenters reduced the need to split their attention between them. Multiple participants (P8, P12, P15, P20, P22, P24, P30) found the \edits{equitable, spatial layout of} \probes \pquote{easier to follow along---see both the presenter and the visualizations}{P15} while the slides version felt like \pquote{two videos were fighting for their attention}{P20}. Even for participants who \pquote{prefer seeing data rather than people [it was] good to see person and data at the same time}{P25}. \edits{Thus, it is not only evident that the new layout changed the relationship dynamics between the visualizations and presenters for viewers, but also, at times, even preferred. }

Furthermore, more than half the participants switched from focusing on the visualizations to the presenter with \probes (\walmartVis 19/30, \injuryVis 17/30, \napoleonVis  17/30, \airplaneVis 16/30; \autoref{fig:breakdown_vis_or_presenter}). These results are non-obvious because many of the visualizations, other than \injuryVis, were not directly overlaid over the presenter at all times---the airplane was detached from the presenter, the French army was at the side and in front of the presenter many of the times, and the map for \walmartVis was located at the torso. 

\edits{At the same time, we also learn from comparing \walmartVis and \injuryVis that \textit{how} we integrate visualizations with the presenters also affect the viewer's experience. We designed \walmartVis as a controlled counterxample to showcae instances of improper integration of presenter and visualization. This was to understand when visualizations should not be combined with the presenter and to assess whether participants would recognize these issues.} 

\edits{As expected,} viewers significantly preferred the baseline \walmartVis over the \probe version in all measures. A significant number of viewers thought the presenter's body movement looked unnatural (\walmartVis 21/30, \autoref{fig:breakdown_slides_or_nonslides}), and for some even \pquote{awkward}{P10}. The map was not only \pquote{distracting}{P20} but also \pquote{not very acceptable when you're trying to learn}{P4}. Consequently, \walmartVis is the only performance in which viewers significantly preferred the baseline over \probes on the metrics of enjoyability, storytelling, and information understanding. However, our results indicate that the merging of spaces did not warrant such negative reactions. Viewers perceived the presenter in \walmartVis as immersed, engaged, and co-existing in the same space as the visualization. \edits{Given that many viewers had positive reactions with \injuryVis, we also know that viewers are not necessarily opposed to the act of attaching visualizations to the body. Thus, the results suggest that the viewers are being affected by \textit{how} the presenter uses the props.} 

These findings indicate a need to develop new tools that help us investigate viewer attention in 3D presentations. Such tools can help presentation designers understand how they can optimize 3D space and orient visualization props \textit{in relation to} presenters to enhance storytelling. Similar to how studies investigate visual attention and flow in static information layouts with eye-tracking~\cite{papoutsaki2017searchgazer,bylinskii2022memorability,borkin2015beyond,lu2020exploring}, more research needs to be conducted to fully understand what components split the viewer attention from the presenter and the visualization when they co-exist on the same 2D screen and 3D space. These visual design patterns can then inform how to create authoring tools that help with strategic blocking to enhance the narrative of the performance.

\paragraph{\edits{\textbf{Merging Spaces Necessitates New Social Engineering Considerations}}}
While most people enjoyed \probes, some did not, as their expectations of presentations did not align with their pre-existing mental models. Some viewers did not like the \napoleonVis \probe because their \pquote{brain associates [presentations] with a more academic setting}{P17}. Participants believed historical data should be presented in the lecture hall settings and not like an InfoVid. P19 believed that for any presentations, \mylongquote{The [presenter] is not essential to understanding information, human connection isn't that necessary.}

\edits{ While we anticipated the new relationship dynamics introduced by \probes would impact the viewers, we did not foresee that these relational dynamics would conflict with the pre-existing ones in the viewers.} Thus, these results indicate that there may first need to be a change in a viewer's mental model and norms of social acceptability associated with presentations on data before such \probes can be fully accepted by a broader audience. 

In addition, we find that visual preferences also affected viewer expectations. The AR elements, while enjoyed by many, for some \pquote{[were] disorienting... clarity of the visual separation [by the slides] made it more immersive as a learning experience}{P21}. While the red bubbles of \injuryVis were effective storytelling devices for most, others preferred the baseline because they thought  \pquote{the red graphic overtook too much of the body}{P26} and thought the presenter looked \pquote{comical when it's moving}{P19}. 
This explains why \injuryVis tied with the baseline on the metric of enjoyability and preferred on the baseline for information understanding. 

Thus, future tools should guide designers on how to design visualizations \textit{in relation} to the presenter, such that they are not too distracting to the performance. These tools should also ensure that the presenter does not compete with the visualization to muddy the message. As P12 points out: 
\mylongquote{Maybe in [certain] situations...you should be aware of the applicability [to] all kinds of people...[if] it focuses more on the presenter themselves and their identity, maybe there could be tension in that.} 

Similar to how character appearance in game design and face filters affect the self-portrayal and identity of individuals in mixed reality~\cite{fribourg2021mirror,morris2023don,birk2013control,chung2023negotiating}, future authoring systems for visual performances should consider the social and visual context that the presenter brings into the performance. Else, mismatches in presentation content and the context the presenter brings will lead to discomfort. \edits{While these were considerations we did not anticipate, the positive reception to the \airplaneVis, \napoleonVis, and \injuryVis \probes indicate that with proper social engineering, \probes offer a unique experience that traditional videoconferencing formats cannot provide.}

\section{Lessons Learned from Desiging \probes: An Autobiographical Perspective}
\label{sec:lessonslearned}
In this section we discuss the lessons learned from the findings and from our nine-month experience designing InfoVids. While we cannot claim generalizability, as Neustaedter and Sengers state~\cite{neustaedter2012autobiographical}, we include insights from an autobiographical standpoint because they provide practical and long-term insights~\cite{huang2023irchiver,zhou2024portalink,desjardins2018revealing} for future visualization presentation tools. 

\paragraph{\textbf{Changing Presenter-System-Viewer Dynamics}}

\begin{figure*}[!htb]
    \centering
    \includegraphics[width=0.8\textwidth]{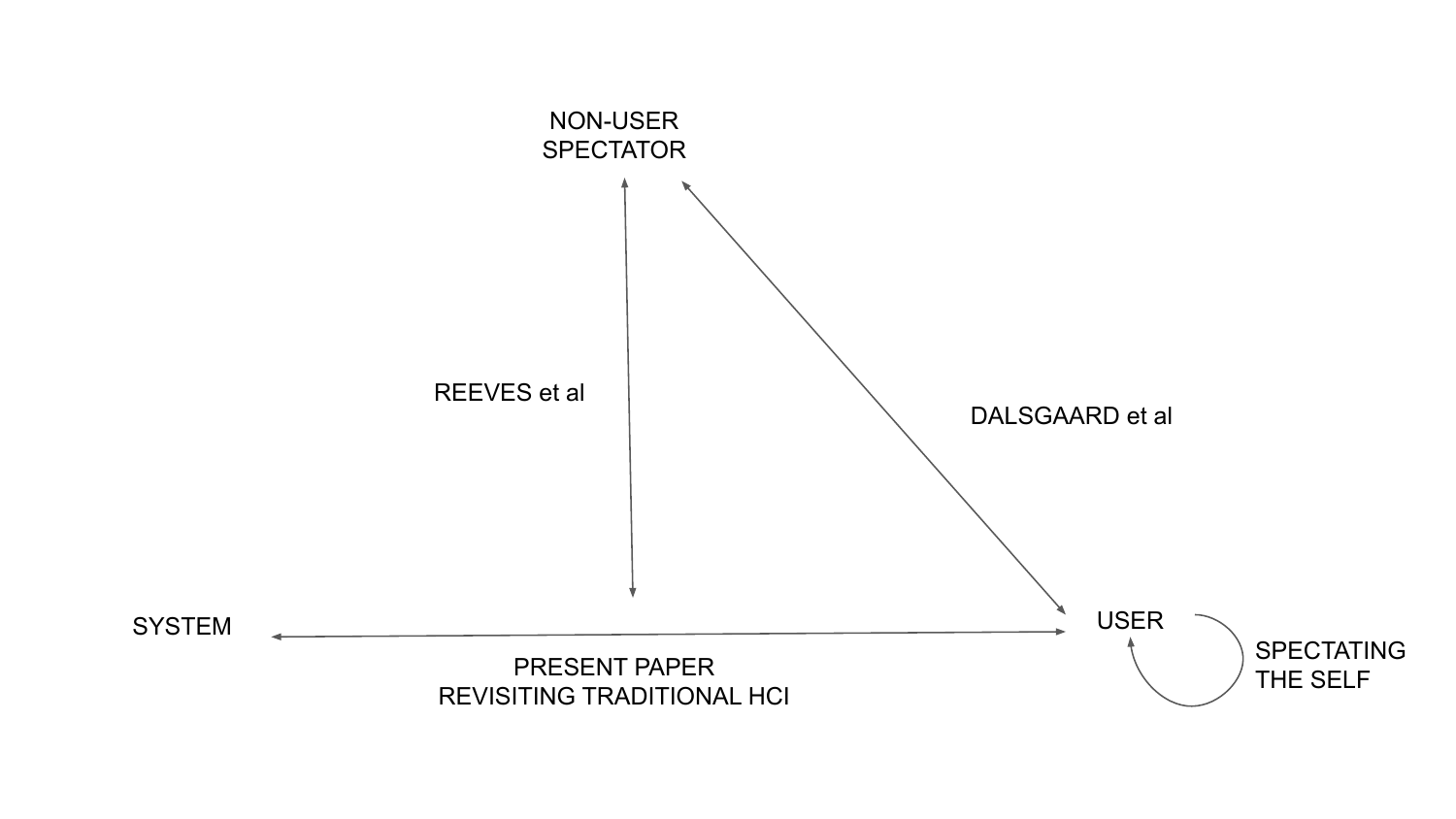}
    \Description{Shows relationship in a triangular for the following three factors: the non-user spectator, system, and the user. Between user and system there’s a double arrow connecting them with the line having the associated label of `present paper, revisiting traditional hci’ to emphasize the focus of the paper. There is a double line connecting the `non-user spectator’ with the `user’ with an associated label `Dalsgaard et al’ to demonstrate what Dalsgaard et al emphasized. Similarly, there’s a double line from `non-user spectator’ to the middle of system-user line with associated label `Reeves et al’. There is also a self-looped arrow for the user with the associated caption `spectating the self’ to emphasize the fact that the user is also a spectator of themselves.}
   \caption{Self-facing cameras make users, or presenters, spectators of themselves. Thus, this paper modifies the original framework proposed by Dalsgaard et al.~\cite{dalsgaard2008performing}  framework by adding a self-looping arrow for the `user' to highlight new user-system-viewer dynamics.}
    \label{fig:dalsgaardModified}
\end{figure*}

\edits{To articulate the experience of spectators in emerging public performances, Reeves et al.~\cite{reeves2005designing} formalizes a classification and taxonomy for performative interfaces. Dalsgaard et al.~\cite{dalsgaard2008performing} later on extends Reeves et al.'s framework to emphasize the user-spectator relationship more than the spectator and user-system relationship and the traditionally-investigated user-system relationships. In spirit, both Reeves et al.~and Dalsgaard et al. describe the emerging role of spectators as active participants affecting the user's performance.}

\edits{However, as we were filming ourselves as the presenter and simultaneously looking at the augmented visualizations overlaid to our body as a viewer, we found that the division between the spectator and the presenter blurred. \edits{The relationship between the presenter and viewer was being redefined.} As viewers, we could both perform \textit{and} view ourselves in real-time with the self-facing camera. This suggests that similar presentation systems found on streaming platforms, such as Twitch, TikTok, and Discord and videoconferencing platforms such as Skype, Teams, and Zoom 
have also introduced new user-system-spectator dynamics. }

\edits{These setups, however, cannot be fully explained by pre-existing frameworks of user-system-spectator relationships~\cite{reeves2005designing, dalsgaard2008performing}. As we demonstrate in \autoref{fig:dalsgaardModified}, we may need to update pre-existing performance theories to account for self-facing cameras and streaming~\cite{lu2018you, lottridge2017third, cheung2011starcraft} which have now muddied the division between the user and the spectator in the opposite direction suggested by Reeves et al. and Dalsgaard et al.~\cite{reeves2005designing, dalsgaard2008performing}. Thus, as virtual and physical spaces start to blend, our probe with \probes prompts the need for future work to investigate the emerging tensions not only between presenter and visualization, but also among the self-spectating presenter, system, and viewer. }

\paragraph{\textbf{Dynamic Presenter-Visualization Relationships}} We also learn that \probes do more than blend space---\probes change the presenter's relationship with the visualization \edits{and influence the message conveyed to the viewers. At times, the presenter blends in with and becomes one with the prop. For example, the presenter's full body acts as the visual highlight in \airplaneVis to indicate where the viewer should look. Presenters are the context and information in \injuryVis and the starter of an engaging visual animation in \napoleonVis.}

\edits{However, these relationships can also be dynamic and evolve over time.} As we used augmented reality to visually place data in real life settings, the data physicalized~\cite{jansen2015opportunities}. However, unlike physical data visualizations, which remain relatively static, the properties of virtual data can be dynamic---they can change in size, location, material, and style. \edits{As a result, the relationship between visualization and presenter can constantly change.} Visualizations that change in size can change how they occupy space, and ultimately alter how a presenter can interact with them. \edits{All of these factors redefines the relationship with the presenter and transforms the types of presentations a viewer can experience~\cite{blocking101}.} 

\edits{On the other hand, such changing affordances can be cognitively tiring to maintain. While the presenter has a view of their augmented self with the visualizations, the visualizations are not tangible like physical objects. The intangible form and changing affordances~\cite{norman2013design} warrant different interactions from the presenter~\cite{brooks1988grasping, kister2017grasp, gong2023affordance}, which means presenters have to memorize how to interact with the visualizations. Even if the presentations were brief, a single presenter had to present in four different InfoVids. This required context switching and the ability to memorize various details and body language for each presentation, all of which can be mentally taxing. }

\edits{Yet, this lesson prompts us to consider how future technologies can be designed to better support extended and multiple presentations---scenarios that are often overlooked in prior works. This could lead to new hidden cue systems customized for the presenter, but hidden from the viewers. For example, in the process of designing \probes, we started to develop a system in which a presenter triggered animations based on the spatial location of the presenter. This freed up the hands, reduced the need to memorize different interactions, and allowed the presenter to command the stage more. All of these components can enhance the viewer's experience, as a less fatigued presenter can positively impact the quality and engagement of the presentation.} \edits{And as we show in our findings, if executed well, \probes can prove to be more engaging than traditional formats for the viewer. }

\section{Conclusion} \label{sec:conc}
We approached our investigation to introduce a more equitable presentation paradigm using \probes from two distinct angles: one from the perspective of the viewers and another from a autobiographical perspective as long-term designers and users of \probes. Our study findings suggest that viewers may find \probes to be a more engaging and preferable format than traditional, 2D presentations, if the relationship of the visualization to the presenter is appropriately considered. In the process of designing and implementing \probes, we learn how spatial layout, form, and interactions affect how viewers perceive presenter-visualization relationships. Our insights into the process of making InfoVids will hopefully inform future data performance systems that nurture a new generation of data presenters who dance with data to tell sophisticated stories \edits{for a broader audience}, carrying on the legacy of Prof.~Hans Rosling.

\bibliographystyle{ACM-Reference-Format}
\bibliography{paper}


\begin{thebibliography}{78}


\ifx \showCODEN    \undefined \def \showCODEN     #1{\unskip}     \fi
\ifx \showDOI      \undefined \def \showDOI       #1{#1}\fi
\ifx \showISBNx    \undefined \def \showISBNx     #1{\unskip}     \fi
\ifx \showISBNxiii \undefined \def \showISBNxiii  #1{\unskip}     \fi
\ifx \showISSN     \undefined \def \showISSN      #1{\unskip}     \fi
\ifx \showLCCN     \undefined \def \showLCCN      #1{\unskip}     \fi
\ifx \shownote     \undefined \def \shownote      #1{#1}          \fi
\ifx \showarticletitle \undefined \def \showarticletitle #1{#1}   \fi
\ifx \showURL      \undefined \def \showURL       {\relax}        \fi
\providecommand\bibfield[2]{#2}
\providecommand\bibinfo[2]{#2}
\providecommand\natexlab[1]{#1}
\providecommand\showeprint[2][]{arXiv:#2}

\bibitem[\protect\citeauthoryear{Amini, Henry~Riche, Lee, Hurter, and Irani}{Amini et~al\mbox{.}}{2015}]%
        {amini2015understanding}
\bibfield{author}{\bibinfo{person}{Fereshteh Amini}, \bibinfo{person}{Nathalie Henry~Riche}, \bibinfo{person}{Bongshin Lee}, \bibinfo{person}{Christophe Hurter}, {and} \bibinfo{person}{Pourang Irani}.} \bibinfo{year}{2015}\natexlab{}.
\newblock \showarticletitle{Understanding data videos: Looking at narrative visualization through the cinematography lens}. In \bibinfo{booktitle}{\emph{Proceedings of the 33rd Annual ACM conference on human factors in computing systems}}. \bibinfo{pages}{1459--1468}.
\newblock


\bibitem[\protect\citeauthoryear{Amini, Riche, Lee, Leboe-McGowan, and Irani}{Amini et~al\mbox{.}}{2018}]%
        {amini2018hooked}
\bibfield{author}{\bibinfo{person}{Fereshteh Amini}, \bibinfo{person}{Nathalie~Henry Riche}, \bibinfo{person}{Bongshin Lee}, \bibinfo{person}{Jason Leboe-McGowan}, {and} \bibinfo{person}{Pourang Irani}.} \bibinfo{year}{2018}\natexlab{}.
\newblock \showarticletitle{Hooked on data videos: assessing the effect of animation and pictographs on viewer engagement}. In \bibinfo{booktitle}{\emph{Proceedings of the 2018 international conference on advanced visual interfaces}}. \bibinfo{pages}{1--9}.
\newblock


\bibitem[\protect\citeauthoryear{Bateman, Mandryk, Gutwin, Genest, McDine, and Brooks}{Bateman et~al\mbox{.}}{2010}]%
        {bateman2010useful}
\bibfield{author}{\bibinfo{person}{Scott Bateman}, \bibinfo{person}{Regan~L Mandryk}, \bibinfo{person}{Carl Gutwin}, \bibinfo{person}{Aaron Genest}, \bibinfo{person}{David McDine}, {and} \bibinfo{person}{Christopher Brooks}.} \bibinfo{year}{2010}\natexlab{}.
\newblock \showarticletitle{Useful junk? The effects of visual embellishment on comprehension and memorability of charts}. In \bibinfo{booktitle}{\emph{Proceedings of the SIGCHI conference on human factors in computing systems}}. \bibinfo{pages}{2573--2582}.
\newblock


\bibitem[\protect\citeauthoryear{Birk and Mandryk}{Birk and Mandryk}{2013}]%
        {birk2013control}
\bibfield{author}{\bibinfo{person}{Max Birk} {and} \bibinfo{person}{Regan~L Mandryk}.} \bibinfo{year}{2013}\natexlab{}.
\newblock \showarticletitle{Control your game-self: effects of controller type on enjoyment, motivation, and personality in game}. In \bibinfo{booktitle}{\emph{Proceedings of the SIGCHI Conference on Human Factors in Computing Systems}}. \bibinfo{pages}{685--694}.
\newblock


\bibitem[\protect\citeauthoryear{Blum, Kleeberger, Bichlmeier, and Navab}{Blum et~al\mbox{.}}{2012}]%
        {blum2012mirracle}
\bibfield{author}{\bibinfo{person}{Tobias Blum}, \bibinfo{person}{Valerie Kleeberger}, \bibinfo{person}{Christoph Bichlmeier}, {and} \bibinfo{person}{Nassir Navab}.} \bibinfo{year}{2012}\natexlab{}.
\newblock \showarticletitle{mirracle: An augmented reality magic mirror system for anatomy education}. In \bibinfo{booktitle}{\emph{2012 IEEE Virtual Reality Workshops (VRW)}}. IEEE, \bibinfo{pages}{115--116}.
\newblock
\urldef\tempurl%
\url{https://doi.org/10.1109/VR.2012.6180909}
\showDOI{\tempurl}


\bibitem[\protect\citeauthoryear{Borkin, Bylinskii, Kim, Bainbridge, Yeh, Borkin, Pfister, and Oliva}{Borkin et~al\mbox{.}}{2015}]%
        {borkin2015beyond}
\bibfield{author}{\bibinfo{person}{Michelle~A Borkin}, \bibinfo{person}{Zoya Bylinskii}, \bibinfo{person}{Nam~Wook Kim}, \bibinfo{person}{Constance~May Bainbridge}, \bibinfo{person}{Chelsea~S Yeh}, \bibinfo{person}{Daniel Borkin}, \bibinfo{person}{Hanspeter Pfister}, {and} \bibinfo{person}{Aude Oliva}.} \bibinfo{year}{2015}\natexlab{}.
\newblock \showarticletitle{Beyond memorability: Visualization recognition and recall}.
\newblock \bibinfo{journal}{\emph{IEEE transactions on visualization and computer graphics}} \bibinfo{volume}{22}, \bibinfo{number}{1} (\bibinfo{year}{2015}), \bibinfo{pages}{519--528}.
\newblock


\bibitem[\protect\citeauthoryear{Bostock}{Bostock}{2023}]%
        {walmartExampleVis}
\bibfield{author}{\bibinfo{person}{Mike Bostock}.} \bibinfo{year}{Accessed in 2023}\natexlab{}.
\newblock \bibinfo{title}{Walmart’s Growth}.
\newblock
\newblock
\urldef\tempurl%
\url{https://observablehq.com/@d3/walmarts-growth}
\showURL{%
\tempurl}


\bibitem[\protect\citeauthoryear{Bostock, Ogievetsky, and Heer}{Bostock et~al\mbox{.}}{2011}]%
        {bostock2011d3}
\bibfield{author}{\bibinfo{person}{Michael Bostock}, \bibinfo{person}{Vadim Ogievetsky}, {and} \bibinfo{person}{Jeffrey Heer}.} \bibinfo{year}{2011}\natexlab{}.
\newblock \showarticletitle{D$^3$ data-driven documents}.
\newblock \bibinfo{journal}{\emph{IEEE transactions on visualization and computer graphics}} \bibinfo{volume}{17}, \bibinfo{number}{12} (\bibinfo{year}{2011}), \bibinfo{pages}{2301--2309}.
\newblock


\bibitem[\protect\citeauthoryear{Braun and Clarke}{Braun and Clarke}{2006}]%
        {braun2006using}
\bibfield{author}{\bibinfo{person}{Virginia Braun} {and} \bibinfo{person}{Victoria Clarke}.} \bibinfo{year}{2006}\natexlab{}.
\newblock \showarticletitle{Using thematic analysis in psychology}.
\newblock \bibinfo{journal}{\emph{Qualitative research in psychology}} \bibinfo{volume}{3}, \bibinfo{number}{2} (\bibinfo{year}{2006}), \bibinfo{pages}{77--101}.
\newblock


\bibitem[\protect\citeauthoryear{Brooks}{Brooks}{1988}]%
        {brooks1988grasping}
\bibfield{author}{\bibinfo{person}{F.~P. Brooks}.} \bibinfo{year}{1988}\natexlab{}.
\newblock \showarticletitle{Grasping Reality through Illusion—interactive Graphics Serving Science}. In \bibinfo{booktitle}{\emph{Proceedings of the SIGCHI Conference on Human Factors in Computing Systems}} (Washington, D.C., USA) \emph{(\bibinfo{series}{CHI '88})}. \bibinfo{pages}{1–11}.
\newblock
\showISBNx{0201142376}
\urldef\tempurl%
\url{https://doi.org/10.1145/57167.57168}
\showDOI{\tempurl}


\bibitem[\protect\citeauthoryear{Buchenau and Suri}{Buchenau and Suri}{2000}]%
        {buchenau2000experience}
\bibfield{author}{\bibinfo{person}{Marion Buchenau} {and} \bibinfo{person}{Jane~Fulton Suri}.} \bibinfo{year}{2000}\natexlab{}.
\newblock \showarticletitle{Experience prototyping}. In \bibinfo{booktitle}{\emph{Proceedings of the 3rd conference on Designing interactive systems: processes, practices, methods, and techniques}}. \bibinfo{pages}{424--433}.
\newblock


\bibitem[\protect\citeauthoryear{Bylinskii, Goetschalckx, Newman, and Oliva}{Bylinskii et~al\mbox{.}}{2022}]%
        {bylinskii2022memorability}
\bibfield{author}{\bibinfo{person}{Zoya Bylinskii}, \bibinfo{person}{Lore Goetschalckx}, \bibinfo{person}{Anelise Newman}, {and} \bibinfo{person}{Aude Oliva}.} \bibinfo{year}{2022}\natexlab{}.
\newblock \bibinfo{booktitle}{\emph{Memorability: An Image-Computable Measure of Information Utility}}.
\newblock \bibinfo{publisher}{Springer International Publishing}, \bibinfo{address}{Cham}, \bibinfo{pages}{207--239}.
\newblock
\showISBNx{978-3-030-81465-6}
\urldef\tempurl%
\url{https://doi.org/10.1007/978-3-030-81465-6_8}
\showDOI{\tempurl}


\bibitem[\protect\citeauthoryear{Chen, Li, Song, and Wang}{Chen et~al\mbox{.}}{2023}]%
        {chen2023iarvis}
\bibfield{author}{\bibinfo{person}{Junjie Chen}, \bibinfo{person}{Chenhui Li}, \bibinfo{person}{Sicheng Song}, {and} \bibinfo{person}{Changbo Wang}.} \bibinfo{year}{2023}\natexlab{}.
\newblock \showarticletitle{iARVis: Mobile AR Based Declarative Information Visualization Authoring, Exploring and Sharing}. In \bibinfo{booktitle}{\emph{2023 IEEE Conference Virtual Reality and 3D User Interfaces (VR)}}. \bibinfo{pages}{11--21}.
\newblock


\bibitem[\protect\citeauthoryear{Chen, Su, Wang, Wang, Qu, and Wu}{Chen et~al\mbox{.}}{2019}]%
        {chen2019marvist}
\bibfield{author}{\bibinfo{person}{Zhutian Chen}, \bibinfo{person}{Yijia Su}, \bibinfo{person}{Yifang Wang}, \bibinfo{person}{Qianwen Wang}, \bibinfo{person}{Huamin Qu}, {and} \bibinfo{person}{Yingcai Wu}.} \bibinfo{year}{2019}\natexlab{}.
\newblock \showarticletitle{Marvist: Authoring glyph-based visualization in mobile augmented reality}.
\newblock \bibinfo{journal}{\emph{IEEE transactions on visualization and computer graphics}} \bibinfo{volume}{26}, \bibinfo{number}{8} (\bibinfo{year}{2019}), \bibinfo{pages}{2645--2658}.
\newblock


\bibitem[\protect\citeauthoryear{Chen, Tong, Wang, Bach, and Qu}{Chen et~al\mbox{.}}{2020}]%
        {chen2020augmenting}
\bibfield{author}{\bibinfo{person}{Zhutian Chen}, \bibinfo{person}{Wai Tong}, \bibinfo{person}{Qianwen Wang}, \bibinfo{person}{Benjamin Bach}, {and} \bibinfo{person}{Huamin Qu}.} \bibinfo{year}{2020}\natexlab{}.
\newblock \showarticletitle{Augmenting static visualizations with paparvis designer}. In \bibinfo{booktitle}{\emph{Proceedings of the 2020 CHI Conference on Human Factors in Computing Systems}}. \bibinfo{pages}{1--12}.
\newblock


\bibitem[\protect\citeauthoryear{Cheung and Huang}{Cheung and Huang}{2011}]%
        {cheung2011starcraft}
\bibfield{author}{\bibinfo{person}{Gifford Cheung} {and} \bibinfo{person}{Jeff Huang}.} \bibinfo{year}{2011}\natexlab{}.
\newblock \showarticletitle{Starcraft from the stands: understanding the game spectator}. In \bibinfo{booktitle}{\emph{Proceedings of the SIGCHI conference on human factors in computing systems}}. \bibinfo{pages}{763--772}.
\newblock


\bibitem[\protect\citeauthoryear{Chung, Fu, Deocadiz-Smith, Jung, and Huang}{Chung et~al\mbox{.}}{2023}]%
        {chung2023negotiating}
\bibfield{author}{\bibinfo{person}{Ji~Won Chung}, \bibinfo{person}{Xiyu~Jenny Fu}, \bibinfo{person}{Zachary Deocadiz-Smith}, \bibinfo{person}{Malte~F Jung}, {and} \bibinfo{person}{Jeff Huang}.} \bibinfo{year}{2023}\natexlab{}.
\newblock \showarticletitle{Negotiating Dyadic Interactions through the Lens of Augmented Reality Glasses}. In \bibinfo{booktitle}{\emph{Proceedings of the 2023 ACM Designing Interactive Systems Conference}}. \bibinfo{pages}{493--508}.
\newblock


\bibitem[\protect\citeauthoryear{Cordeil, Cunningham, Bach, Hurter, Thomas, Marriott, and Dwyer}{Cordeil et~al\mbox{.}}{2019}]%
        {cordeil2019iatk}
\bibfield{author}{\bibinfo{person}{Maxime Cordeil}, \bibinfo{person}{Andrew Cunningham}, \bibinfo{person}{Benjamin Bach}, \bibinfo{person}{Christophe Hurter}, \bibinfo{person}{Bruce~H Thomas}, \bibinfo{person}{Kim Marriott}, {and} \bibinfo{person}{Tim Dwyer}.} \bibinfo{year}{2019}\natexlab{}.
\newblock \showarticletitle{IATK: An immersive analytics toolkit}. In \bibinfo{booktitle}{\emph{2019 IEEE Conference on Virtual Reality and 3D User Interfaces (VR)}}. IEEE, \bibinfo{pages}{200--209}.
\newblock


\bibitem[\protect\citeauthoryear{Dalsgaard and Hansen}{Dalsgaard and Hansen}{2008}]%
        {dalsgaard2008performing}
\bibfield{author}{\bibinfo{person}{Peter Dalsgaard} {and} \bibinfo{person}{Lone~Koefoed Hansen}.} \bibinfo{year}{2008}\natexlab{}.
\newblock \showarticletitle{Performing perception—staging aesthetics of interaction}.
\newblock \bibinfo{journal}{\emph{ACM Transactions on Computer-Human Interaction (TOCHI)}} \bibinfo{volume}{15}, \bibinfo{number}{3} (\bibinfo{year}{2008}), \bibinfo{pages}{1--33}.
\newblock


\bibitem[\protect\citeauthoryear{Davis, Asente, and Yang}{Davis et~al\mbox{.}}{2023}]%
        {davis2023multimodal}
\bibfield{author}{\bibinfo{person}{Josh~Urban Davis}, \bibinfo{person}{Paul Asente}, {and} \bibinfo{person}{Xing-Dong Yang}.} \bibinfo{year}{2023}\natexlab{}.
\newblock \showarticletitle{Multimodal Direct Manipulation in Video Conferencing: Challenges and Opportunities}. In \bibinfo{booktitle}{\emph{Proceedings of the 2023 ACM Designing Interactive Systems Conference}}. \bibinfo{pages}{1174--1193}.
\newblock


\bibitem[\protect\citeauthoryear{Denning, Dehlawi, and Kohno}{Denning et~al\mbox{.}}{2014}]%
        {denning2014situ}
\bibfield{author}{\bibinfo{person}{Tamara Denning}, \bibinfo{person}{Zakariya Dehlawi}, {and} \bibinfo{person}{Tadayoshi Kohno}.} \bibinfo{year}{2014}\natexlab{}.
\newblock \showarticletitle{In situ with bystanders of augmented reality glasses: Perspectives on recording and privacy-mediating technologies}. In \bibinfo{booktitle}{\emph{Proceedings of the SIGCHI Conference on Human Factors in Computing Systems}}. \bibinfo{pages}{2377--2386}.
\newblock
\urldef\tempurl%
\url{https://doi.org/10.1145/2556288.2557352}
\showURL{%
\tempurl}


\bibitem[\protect\citeauthoryear{Desjardins and Ball}{Desjardins and Ball}{2018}]%
        {desjardins2018revealing}
\bibfield{author}{\bibinfo{person}{Audrey Desjardins} {and} \bibinfo{person}{Aubree Ball}.} \bibinfo{year}{2018}\natexlab{}.
\newblock \showarticletitle{Revealing tensions in autobiographical design in HCI}. In \bibinfo{booktitle}{\emph{proceedings of the 2018 designing interactive systems conference}}. \bibinfo{pages}{753--764}.
\newblock


\bibitem[\protect\citeauthoryear{Fribourg, Peillard, and Mcdonnell}{Fribourg et~al\mbox{.}}{2021}]%
        {fribourg2021mirror}
\bibfield{author}{\bibinfo{person}{Rebecca Fribourg}, \bibinfo{person}{Etienne Peillard}, {and} \bibinfo{person}{Rachel Mcdonnell}.} \bibinfo{year}{2021}\natexlab{}.
\newblock \showarticletitle{Mirror, mirror on my phone: Investigating dimensions of self-face perception induced by augmented reality filters}. In \bibinfo{booktitle}{\emph{2021 IEEE International Symposium on Mixed and Augmented Reality (ISMAR)}}. IEEE, \bibinfo{pages}{470--478}.
\newblock


\bibitem[\protect\citeauthoryear{Friendly}{Friendly}{2023}]%
        {minardExampleVis}
\bibfield{author}{\bibinfo{person}{Michael Friendly}.} \bibinfo{year}{Accessed in 2023}\natexlab{}.
\newblock \bibinfo{title}{Minard's Graphic Works}.
\newblock
\newblock
\urldef\tempurl%
\url{https://www.datavis.ca/gallery/re-minard.php}
\showURL{%
\tempurl}


\bibitem[\protect\citeauthoryear{Gong, Santosa, Grossman, Glueck, Clarke, and Lai}{Gong et~al\mbox{.}}{2023}]%
        {gong2023affordance}
\bibfield{author}{\bibinfo{person}{Weilun Gong}, \bibinfo{person}{Stephanie Santosa}, \bibinfo{person}{Tovi Grossman}, \bibinfo{person}{Michael Glueck}, \bibinfo{person}{Daniel Clarke}, {and} \bibinfo{person}{Frances Lai}.} \bibinfo{year}{2023}\natexlab{}.
\newblock \showarticletitle{Affordance-Based and User-Defined Gestures for Spatial Tangible Interaction}. In \bibinfo{booktitle}{\emph{Proceedings of the 2023 ACM Designing Interactive Systems Conference}}. \bibinfo{pages}{1500--1514}.
\newblock


\bibitem[\protect\citeauthoryear{Graham}{Graham}{2024}]%
        {planeExampleVis}
\bibfield{author}{\bibinfo{person}{Tim Graham}.} \bibinfo{year}{Accessed in 2024}\natexlab{}.
\newblock \bibinfo{title}{The Safest Seat to Sit In On a Plane is…}.
\newblock
\newblock
\urldef\tempurl%
\url{https://flowingdata.com/2008/05/20/the-safest-seat-to-sit-in-on-a-plane-is/}
\showURL{%
\tempurl}


\bibitem[\protect\citeauthoryear{Grime}{Grime}{2015}]%
        {minardNumberphile}
\bibfield{author}{\bibinfo{person}{James Grime}.} \bibinfo{year}{2015}\natexlab{}.
\newblock \bibinfo{title}{The Greatest Ever Infographic - Numberphile}.
\newblock
\newblock
\urldef\tempurl%
\url{https://www.youtube.com/watch?v=3T7jMcstxY0&ab_channel=Numberphile}
\showURL{%
\tempurl}


\bibitem[\protect\citeauthoryear{Hall, Bartram, and Brehmer}{Hall et~al\mbox{.}}{2022}]%
        {hall2022augmented}
\bibfield{author}{\bibinfo{person}{Brian~D Hall}, \bibinfo{person}{Lyn Bartram}, {and} \bibinfo{person}{Matthew Brehmer}.} \bibinfo{year}{2022}\natexlab{}.
\newblock \showarticletitle{Augmented chironomia for presenting data to remote audiences}. In \bibinfo{booktitle}{\emph{Proceedings of the 35th Annual ACM Symposium on User Interface Software and Technology}}. \bibinfo{pages}{1--14}.
\newblock


\bibitem[\protect\citeauthoryear{Harrison, Reinecke, and Chang}{Harrison et~al\mbox{.}}{2015}]%
        {harrison2015infographic}
\bibfield{author}{\bibinfo{person}{Lane Harrison}, \bibinfo{person}{Katharina Reinecke}, {and} \bibinfo{person}{Remco Chang}.} \bibinfo{year}{2015}\natexlab{}.
\newblock \showarticletitle{Infographic aesthetics: Designing for the first impression}. In \bibinfo{booktitle}{\emph{Proceedings of the 33rd Annual ACM conference on human factors in computing systems}}. \bibinfo{pages}{1187--1190}.
\newblock


\bibitem[\protect\citeauthoryear{Hassoun, Beacock, Consolvo, Goldberg, Kelley, and Russell}{Hassoun et~al\mbox{.}}{2023}]%
        {hassoun2023practicing}
\bibfield{author}{\bibinfo{person}{Amelia Hassoun}, \bibinfo{person}{Ian Beacock}, \bibinfo{person}{Sunny Consolvo}, \bibinfo{person}{Beth Goldberg}, \bibinfo{person}{Patrick~Gage Kelley}, {and} \bibinfo{person}{Daniel~M Russell}.} \bibinfo{year}{2023}\natexlab{}.
\newblock \showarticletitle{Practicing Information Sensibility: How Gen Z Engages with Online Information}. In \bibinfo{booktitle}{\emph{Proceedings of the 2023 CHI Conference on Human Factors in Computing Systems}}. \bibinfo{pages}{1--17}.
\newblock


\bibitem[\protect\citeauthoryear{Huang and Qian}{Huang and Qian}{2023}]%
        {huang2023irchiver}
\bibfield{author}{\bibinfo{person}{Jeff Huang} {and} \bibinfo{person}{Jing Qian}.} \bibinfo{year}{2023}\natexlab{}.
\newblock \showarticletitle{irchiver: A Full-Resolution Personal Web Archive for Users and Researchers}. In \bibinfo{booktitle}{\emph{Proceedings of the 2023 Conference on Human Information Interaction and Retrieval}}. \bibinfo{pages}{449--453}.
\newblock


\bibitem[\protect\citeauthoryear{Hubenschmid, Zagermann, Butscher, and Reiterer}{Hubenschmid et~al\mbox{.}}{2021}]%
        {hubenschmid2021stream}
\bibfield{author}{\bibinfo{person}{Sebastian Hubenschmid}, \bibinfo{person}{Johannes Zagermann}, \bibinfo{person}{Simon Butscher}, {and} \bibinfo{person}{Harald Reiterer}.} \bibinfo{year}{2021}\natexlab{}.
\newblock \showarticletitle{Stream: Exploring the combination of spatially-aware tablets with augmented reality head-mounted displays for immersive analytics}. In \bibinfo{booktitle}{\emph{Proceedings of the 2021 CHI Conference on Human Factors in Computing Systems}}. \bibinfo{pages}{1--14}.
\newblock


\bibitem[\protect\citeauthoryear{Hutchinson, Mackay, Westerlund, Bederson, Druin, Plaisant, Beaudouin-Lafon, Conversy, Evans, Hansen, et~al\mbox{.}}{Hutchinson et~al\mbox{.}}{2003}]%
        {hutchinson2003technology}
\bibfield{author}{\bibinfo{person}{Hilary Hutchinson}, \bibinfo{person}{Wendy Mackay}, \bibinfo{person}{Bo Westerlund}, \bibinfo{person}{Benjamin~B Bederson}, \bibinfo{person}{Allison Druin}, \bibinfo{person}{Catherine Plaisant}, \bibinfo{person}{Michel Beaudouin-Lafon}, \bibinfo{person}{St{\'e}phane Conversy}, \bibinfo{person}{Helen Evans}, \bibinfo{person}{Heiko Hansen}, {et~al\mbox{.}}} \bibinfo{year}{2003}\natexlab{}.
\newblock \showarticletitle{Technology probes: inspiring design for and with families}. In \bibinfo{booktitle}{\emph{Proceedings of the SIGCHI conference on Human factors in computing systems}}. \bibinfo{pages}{17--24}.
\newblock


\bibitem[\protect\citeauthoryear{Jansen, Dragicevic, Isenberg, Alexander, Karnik, Kildal, Subramanian, and Hornb{\ae}k}{Jansen et~al\mbox{.}}{2015}]%
        {jansen2015opportunities}
\bibfield{author}{\bibinfo{person}{Yvonne Jansen}, \bibinfo{person}{Pierre Dragicevic}, \bibinfo{person}{Petra Isenberg}, \bibinfo{person}{Jason Alexander}, \bibinfo{person}{Abhijit Karnik}, \bibinfo{person}{Johan Kildal}, \bibinfo{person}{Sriram Subramanian}, {and} \bibinfo{person}{Kasper Hornb{\ae}k}.} \bibinfo{year}{2015}\natexlab{}.
\newblock \showarticletitle{Opportunities and challenges for data physicalization}. In \bibinfo{booktitle}{\emph{proceedings of the 33rd annual acm conference on human factors in computing systems}}. \bibinfo{pages}{3227--3236}.
\newblock


\bibitem[\protect\citeauthoryear{Jiang, Li, He, Lindlbauer, and Yan}{Jiang et~al\mbox{.}}{2023}]%
        {jiang2023handavatar}
\bibfield{author}{\bibinfo{person}{Yu Jiang}, \bibinfo{person}{Zhipeng Li}, \bibinfo{person}{Mufei He}, \bibinfo{person}{David Lindlbauer}, {and} \bibinfo{person}{Yukang Yan}.} \bibinfo{year}{2023}\natexlab{}.
\newblock \showarticletitle{HandAvatar: Embodying Non-Humanoid Virtual Avatars through Hands}. In \bibinfo{booktitle}{\emph{Proceedings of the 2023 CHI Conference on Human Factors in Computing Systems}}. \bibinfo{pages}{1--17}.
\newblock


\bibitem[\protect\citeauthoryear{Kister, Klamka, Tominski, and Dachselt}{Kister et~al\mbox{.}}{2017}]%
        {kister2017grasp}
\bibfield{author}{\bibinfo{person}{Ulrike Kister}, \bibinfo{person}{Konstantin Klamka}, \bibinfo{person}{Christian Tominski}, {and} \bibinfo{person}{Raimund Dachselt}.} \bibinfo{year}{2017}\natexlab{}.
\newblock \showarticletitle{GraSp: Combining Spatially-aware Mobile Devices and a Display Wall for Graph Visualization and Interaction}.
\newblock \bibinfo{journal}{\emph{Computer Graphics Forum}} \bibinfo{volume}{36}, \bibinfo{number}{3} (\bibinfo{year}{2017}), \bibinfo{pages}{503--514}.
\newblock


\bibitem[\protect\citeauthoryear{Kosara and Mackinlay}{Kosara and Mackinlay}{2013}]%
        {kosara2013storytelling}
\bibfield{author}{\bibinfo{person}{Robert Kosara} {and} \bibinfo{person}{Jock Mackinlay}.} \bibinfo{year}{2013}\natexlab{}.
\newblock \showarticletitle{Storytelling: The Next Step for Visualization}.
\newblock \bibinfo{journal}{\emph{Computer}} \bibinfo{volume}{46}, \bibinfo{number}{5} (\bibinfo{year}{2013}), \bibinfo{pages}{44--50}.
\newblock
\urldef\tempurl%
\url{https://doi.org/10.1109/MC.2013.36}
\showDOI{\tempurl}


\bibitem[\protect\citeauthoryear{Langner, Satkowski, B{\"u}schel, and Dachselt}{Langner et~al\mbox{.}}{2021}]%
        {langner2021marvis}
\bibfield{author}{\bibinfo{person}{Ricardo Langner}, \bibinfo{person}{Marc Satkowski}, \bibinfo{person}{Wolfgang B{\"u}schel}, {and} \bibinfo{person}{Raimund Dachselt}.} \bibinfo{year}{2021}\natexlab{}.
\newblock \showarticletitle{Marvis: Combining mobile devices and augmented reality for visual data analysis}. In \bibinfo{booktitle}{\emph{Proceedings of the 2021 CHI Conference on Human Factors in Computing Systems}}. \bibinfo{pages}{1--17}.
\newblock


\bibitem[\protect\citeauthoryear{Lee, Kazi, and Smith}{Lee et~al\mbox{.}}{2013}]%
        {lee2013sketchstory}
\bibfield{author}{\bibinfo{person}{Bongshin Lee}, \bibinfo{person}{Rubaiat~Habib Kazi}, {and} \bibinfo{person}{Greg Smith}.} \bibinfo{year}{2013}\natexlab{}.
\newblock \showarticletitle{SketchStory: Telling more engaging stories with data through freeform sketching}.
\newblock \bibinfo{journal}{\emph{IEEE transactions on visualization and computer graphics}} \bibinfo{volume}{19}, \bibinfo{number}{12} (\bibinfo{year}{2013}), \bibinfo{pages}{2416--2425}.
\newblock


\bibitem[\protect\citeauthoryear{Li and Moacdieh}{Li and Moacdieh}{2014}]%
        {li2014chart}
\bibfield{author}{\bibinfo{person}{Huiyang Li} {and} \bibinfo{person}{Nadine Moacdieh}.} \bibinfo{year}{2014}\natexlab{}.
\newblock \showarticletitle{Is “chart junk” useful? An extended examination of visual embellishment}. In \bibinfo{booktitle}{\emph{Proceedings of the Human Factors and Ergonomics Society Annual Meeting}}, Vol.~\bibinfo{volume}{58}. Sage Publications Sage CA: Los Angeles, CA, \bibinfo{pages}{1516--1520}.
\newblock


\bibitem[\protect\citeauthoryear{Liao, Karim, Jadon, Kazi, and Suzuki}{Liao et~al\mbox{.}}{2022}]%
        {liao2022realitytalk}
\bibfield{author}{\bibinfo{person}{Jian Liao}, \bibinfo{person}{Adnan Karim}, \bibinfo{person}{Shivesh~Singh Jadon}, \bibinfo{person}{Rubaiat~Habib Kazi}, {and} \bibinfo{person}{Ryo Suzuki}.} \bibinfo{year}{2022}\natexlab{}.
\newblock \showarticletitle{RealityTalk: Real-Time Speech-Driven Augmented Presentation for AR Live Storytelling}. In \bibinfo{booktitle}{\emph{Proceedings of the 35th Annual ACM Symposium on User Interface Software and Technology}}. \bibinfo{pages}{1--12}.
\newblock


\bibitem[\protect\citeauthoryear{Lo, Gupta, Shigyo, Wu, Bertini, and Qu}{Lo et~al\mbox{.}}{2022}]%
        {lo2022misinformed}
\bibfield{author}{\bibinfo{person}{Leo Yu-Ho Lo}, \bibinfo{person}{Ayush Gupta}, \bibinfo{person}{Kento Shigyo}, \bibinfo{person}{Aoyu Wu}, \bibinfo{person}{Enrico Bertini}, {and} \bibinfo{person}{Huamin Qu}.} \bibinfo{year}{2022}\natexlab{}.
\newblock \showarticletitle{Misinformed by visualization: What do we learn from misinformative visualizations?}
\newblock \bibinfo{journal}{\emph{Computer Graphics Forum}} \bibinfo{volume}{41}, \bibinfo{number}{3} (\bibinfo{year}{2022}), \bibinfo{pages}{515--525}.
\newblock


\bibitem[\protect\citeauthoryear{Lottridge, Bentley, Wheeler, Lee, Cheung, Ong, and Rowley}{Lottridge et~al\mbox{.}}{2017}]%
        {lottridge2017third}
\bibfield{author}{\bibinfo{person}{Danielle Lottridge}, \bibinfo{person}{Frank Bentley}, \bibinfo{person}{Matt Wheeler}, \bibinfo{person}{Jason Lee}, \bibinfo{person}{Janet Cheung}, \bibinfo{person}{Katherine Ong}, {and} \bibinfo{person}{Cristy Rowley}.} \bibinfo{year}{2017}\natexlab{}.
\newblock \showarticletitle{Third-wave livestreaming: teens' long form selfie}. In \bibinfo{booktitle}{\emph{Proceedings of the 19th international conference on human-computer interaction with mobile devices and services}}. \bibinfo{pages}{1--12}.
\newblock


\bibitem[\protect\citeauthoryear{Lu, Wang, Lanir, Zhao, Pfister, Cohen-Or, and Huang}{Lu et~al\mbox{.}}{2020}]%
        {lu2020exploring}
\bibfield{author}{\bibinfo{person}{Min Lu}, \bibinfo{person}{Chufeng Wang}, \bibinfo{person}{Joel Lanir}, \bibinfo{person}{Nanxuan Zhao}, \bibinfo{person}{Hanspeter Pfister}, \bibinfo{person}{Daniel Cohen-Or}, {and} \bibinfo{person}{Hui Huang}.} \bibinfo{year}{2020}\natexlab{}.
\newblock \showarticletitle{Exploring Visual Information Flows in Infographics}. In \bibinfo{booktitle}{\emph{Proceedings of the 2020 CHI Conference on Human Factors in Computing Systems}} (Honolulu, HI, USA) \emph{(\bibinfo{series}{CHI '20})}. \bibinfo{pages}{1–12}.
\newblock
\showISBNx{9781450367080}
\urldef\tempurl%
\url{https://doi.org/10.1145/3313831.3376263}
\showDOI{\tempurl}


\bibitem[\protect\citeauthoryear{Lu, Xia, Heo, and Wigdor}{Lu et~al\mbox{.}}{2018}]%
        {lu2018you}
\bibfield{author}{\bibinfo{person}{Zhicong Lu}, \bibinfo{person}{Haijun Xia}, \bibinfo{person}{Seongkook Heo}, {and} \bibinfo{person}{Daniel Wigdor}.} \bibinfo{year}{2018}\natexlab{}.
\newblock \showarticletitle{You watch, you give, and you engage: a study of live streaming practices in China}. In \bibinfo{booktitle}{\emph{Proceedings of the 2018 CHI conference on human factors in computing systems}}. \bibinfo{pages}{1--13}.
\newblock


\bibitem[\protect\citeauthoryear{Luo, Goebel, Reipschl{\"a}ger, Ellenberg, and Dachselt}{Luo et~al\mbox{.}}{2021}]%
        {luo2021exploring}
\bibfield{author}{\bibinfo{person}{Weizhou Luo}, \bibinfo{person}{Eva Goebel}, \bibinfo{person}{Patrick Reipschl{\"a}ger}, \bibinfo{person}{Mats~Ole Ellenberg}, {and} \bibinfo{person}{Raimund Dachselt}.} \bibinfo{year}{2021}\natexlab{}.
\newblock \showarticletitle{Exploring and slicing volumetric medical data in augmented reality using a spatially-aware mobile device}. In \bibinfo{booktitle}{\emph{2021 IEEE International Symposium on Mixed and Augmented Reality Adjunct (ISMAR-Adjunct)}}. IEEE, \bibinfo{pages}{334--339}.
\newblock


\bibitem[\protect\citeauthoryear{Luo and Tang}{Luo and Tang}{2008}]%
        {luo2008photo}
\bibfield{author}{\bibinfo{person}{Yiwen Luo} {and} \bibinfo{person}{Xiaoou Tang}.} \bibinfo{year}{2008}\natexlab{}.
\newblock \showarticletitle{Photo and video quality evaluation: Focusing on the subject}. In \bibinfo{booktitle}{\emph{Computer Vision--ECCV 2008: 10th European Conference on Computer Vision, Marseille, France, October 12-18, 2008, Proceedings, Part III 10}}. Springer, \bibinfo{pages}{386--399}.
\newblock


\bibitem[\protect\citeauthoryear{Morris, Rosner, Nurius, and Dolev}{Morris et~al\mbox{.}}{2023}]%
        {morris2023don}
\bibfield{author}{\bibinfo{person}{Margaret~E Morris}, \bibinfo{person}{Daniela~K Rosner}, \bibinfo{person}{Paula~S Nurius}, {and} \bibinfo{person}{Hadar~M Dolev}.} \bibinfo{year}{2023}\natexlab{}.
\newblock \showarticletitle{“I Don't Want to Hide Behind an Avatar”: Self-Representation in Social VR Among Women in Midlife}. In \bibinfo{booktitle}{\emph{Proceedings of the 2023 ACM Designing Interactive Systems Conference}} (Pittsburgh, PA, USA) \emph{(\bibinfo{series}{DIS '23})}. \bibinfo{pages}{537–546}.
\newblock
\showISBNx{9781450398930}
\urldef\tempurl%
\url{https://doi.org/10.1145/3563657.3596129}
\showDOI{\tempurl}


\bibitem[\protect\citeauthoryear{Network}{Network}{2020}]%
        {injuriesExampleVis}
\bibfield{author}{\bibinfo{person}{The~Learning Network}.} \bibinfo{year}{2020}\natexlab{}.
\newblock \bibinfo{title}{What’s going on in this graph? | high-school sports injuries}.
\newblock
\newblock
\urldef\tempurl%
\url{https://www.nytimes.com/2020/01/23/learning/whats-going-on-in-this-graph-high-school-sports-injuries.html}
\showURL{%
\tempurl}


\bibitem[\protect\citeauthoryear{Neustaedter and Sengers}{Neustaedter and Sengers}{2012}]%
        {neustaedter2012autobiographical}
\bibfield{author}{\bibinfo{person}{Carman Neustaedter} {and} \bibinfo{person}{Phoebe Sengers}.} \bibinfo{year}{2012}\natexlab{}.
\newblock \showarticletitle{Autobiographical design in HCI research: designing and learning through use-it-yourself}. In \bibinfo{booktitle}{\emph{Proceedings of the Designing Interactive Systems Conference}}. \bibinfo{pages}{514--523}.
\newblock


\bibitem[\protect\citeauthoryear{Norman}{Norman}{2013}]%
        {norman2013design}
\bibfield{author}{\bibinfo{person}{Don Norman}.} \bibinfo{year}{2013}\natexlab{}.
\newblock \bibinfo{booktitle}{\emph{The design of everyday things: Revised and expanded edition}}.
\newblock \bibinfo{publisher}{Basic books}, \bibinfo{address}{New York, New York}.
\newblock


\bibitem[\protect\citeauthoryear{Norooz, Mauriello, Jorgensen, McNally, and Froehlich}{Norooz et~al\mbox{.}}{2015}]%
        {norooz2015bodyvis}
\bibfield{author}{\bibinfo{person}{Leyla Norooz}, \bibinfo{person}{Matthew~Louis Mauriello}, \bibinfo{person}{Anita Jorgensen}, \bibinfo{person}{Brenna McNally}, {and} \bibinfo{person}{Jon~E Froehlich}.} \bibinfo{year}{2015}\natexlab{}.
\newblock \showarticletitle{BodyVis: A new approach to body learning through wearable sensing and visualization}. In \bibinfo{booktitle}{\emph{Proceedings of the 33rd Annual ACM Conference on Human Factors in Computing Systems}}. \bibinfo{pages}{1025--1034}.
\newblock


\bibitem[\protect\citeauthoryear{Papoutsaki, Laskey, and Huang}{Papoutsaki et~al\mbox{.}}{2017}]%
        {papoutsaki2017searchgazer}
\bibfield{author}{\bibinfo{person}{Alexandra Papoutsaki}, \bibinfo{person}{James Laskey}, {and} \bibinfo{person}{Jeff Huang}.} \bibinfo{year}{2017}\natexlab{}.
\newblock \showarticletitle{Searchgazer: Webcam eye tracking for remote studies of web search}. In \bibinfo{booktitle}{\emph{Proceedings of the 2017 conference on conference human information interaction and retrieval}}. \bibinfo{pages}{17--26}.
\newblock


\bibitem[\protect\citeauthoryear{Pei, Chen, Lee, and Zhang}{Pei et~al\mbox{.}}{2022}]%
        {pei2022hand}
\bibfield{author}{\bibinfo{person}{Siyou Pei}, \bibinfo{person}{Alexander Chen}, \bibinfo{person}{Jaewook Lee}, {and} \bibinfo{person}{Yang Zhang}.} \bibinfo{year}{2022}\natexlab{}.
\newblock \showarticletitle{Hand interfaces: Using hands to imitate objects in AR/VR for expressive interactions}. In \bibinfo{booktitle}{\emph{Proceedings of the 2022 CHI conference on human factors in computing systems}}. \bibinfo{pages}{1--16}.
\newblock


\bibitem[\protect\citeauthoryear{Perlin, He, and Rosenberg}{Perlin et~al\mbox{.}}{2018}]%
        {perlin2018chalktalk}
\bibfield{author}{\bibinfo{person}{Ken Perlin}, \bibinfo{person}{Zhenyi He}, {and} \bibinfo{person}{Karl Rosenberg}.} \bibinfo{year}{2018}\natexlab{}.
\newblock \bibinfo{title}{Chalktalk: A Visualization and Communication Language--As a Tool in the Domain of Computer Science Education}.
\newblock
\newblock
\showeprint[arxiv]{1809.07166}~[cs.HC]


\bibitem[\protect\citeauthoryear{Rajaram and Nebeling}{Rajaram and Nebeling}{2022}]%
        {rajaram2022paper}
\bibfield{author}{\bibinfo{person}{Shwetha Rajaram} {and} \bibinfo{person}{Michael Nebeling}.} \bibinfo{year}{2022}\natexlab{}.
\newblock \showarticletitle{Paper trail: An immersive authoring system for augmented reality instructional experiences}. In \bibinfo{booktitle}{\emph{Proceedings of the 2022 CHI Conference on Human Factors in Computing Systems}}. \bibinfo{pages}{1--16}.
\newblock


\bibitem[\protect\citeauthoryear{Reeves, Benford, O'Malley, and Fraser}{Reeves et~al\mbox{.}}{2005}]%
        {reeves2005designing}
\bibfield{author}{\bibinfo{person}{Stuart Reeves}, \bibinfo{person}{Steve Benford}, \bibinfo{person}{Claire O'Malley}, {and} \bibinfo{person}{Mike Fraser}.} \bibinfo{year}{2005}\natexlab{}.
\newblock \showarticletitle{Designing the spectator experience}. In \bibinfo{booktitle}{\emph{Proceedings of the SIGCHI conference on Human factors in computing systems}}. \bibinfo{pages}{741--750}.
\newblock


\bibitem[\protect\citeauthoryear{Rolsing}{Rolsing}{2011}]%
        {hansRosling}
\bibfield{author}{\bibinfo{person}{Hans Rolsing}.} \bibinfo{year}{2011}\natexlab{}.
\newblock \bibinfo{title}{Hans Rosling's 200 Countries, 200 Years, 4 Minutes}.
\newblock
\newblock
\urldef\tempurl%
\url{https://www.youtube.com/watch?v=jbkSRLYSojo&t=54s&ab_channel=BBC}
\showURL{%
\tempurl}


\bibitem[\protect\citeauthoryear{Sallam, Sakamoto, Leboe-McGowan, Latulipe, and Irani}{Sallam et~al\mbox{.}}{2022}]%
        {sallam2022towards}
\bibfield{author}{\bibinfo{person}{Samar Sallam}, \bibinfo{person}{Yumiko Sakamoto}, \bibinfo{person}{Jason Leboe-McGowan}, \bibinfo{person}{Celine Latulipe}, {and} \bibinfo{person}{Pourang Irani}.} \bibinfo{year}{2022}\natexlab{}.
\newblock \showarticletitle{Towards design guidelines for effective health-related data videos: An empirical investigation of affect, personality, and video content}. In \bibinfo{booktitle}{\emph{Proceedings of the 2022 CHI Conference on Human Factors in Computing Systems}}. \bibinfo{pages}{1--22}.
\newblock


\bibitem[\protect\citeauthoryear{Saquib, Kazi, Wei, and Li}{Saquib et~al\mbox{.}}{2019}]%
        {saquib2019interactive}
\bibfield{author}{\bibinfo{person}{Nazmus Saquib}, \bibinfo{person}{Rubaiat~Habib Kazi}, \bibinfo{person}{Li-Yi Wei}, {and} \bibinfo{person}{Wilmot Li}.} \bibinfo{year}{2019}\natexlab{}.
\newblock \showarticletitle{Interactive body-driven graphics for augmented video performance}. In \bibinfo{booktitle}{\emph{Proceedings of the 2019 CHI Conference on Human Factors in Computing Systems}}. \bibinfo{pages}{1--12}.
\newblock


\bibitem[\protect\citeauthoryear{Satriadi, Smiley, Ens, Cordeil, Czauderna, Lee, Yang, Dwyer, and Jenny}{Satriadi et~al\mbox{.}}{2022}]%
        {satriadi2022tangible}
\bibfield{author}{\bibinfo{person}{Kadek~Ananta Satriadi}, \bibinfo{person}{Jim Smiley}, \bibinfo{person}{Barrett Ens}, \bibinfo{person}{Maxime Cordeil}, \bibinfo{person}{Tobias Czauderna}, \bibinfo{person}{Benjamin Lee}, \bibinfo{person}{Ying Yang}, \bibinfo{person}{Tim Dwyer}, {and} \bibinfo{person}{Bernhard Jenny}.} \bibinfo{year}{2022}\natexlab{}.
\newblock \showarticletitle{Tangible globes for data visualisation in augmented reality}. In \bibinfo{booktitle}{\emph{Proceedings of the 2022 CHI Conference on Human Factors in Computing Systems}}. \bibinfo{pages}{1--16}.
\newblock


\bibitem[\protect\citeauthoryear{Satyanarayan, Moritz, Wongsuphasawat, and Heer}{Satyanarayan et~al\mbox{.}}{2016}]%
        {satyanarayan2016vega}
\bibfield{author}{\bibinfo{person}{Arvind Satyanarayan}, \bibinfo{person}{Dominik Moritz}, \bibinfo{person}{Kanit Wongsuphasawat}, {and} \bibinfo{person}{Jeffrey Heer}.} \bibinfo{year}{2016}\natexlab{}.
\newblock \showarticletitle{Vega-lite: A grammar of interactive graphics}.
\newblock \bibinfo{journal}{\emph{IEEE transactions on visualization and computer graphics}} \bibinfo{volume}{23}, \bibinfo{number}{1} (\bibinfo{year}{2016}), \bibinfo{pages}{341--350}.
\newblock


\bibitem[\protect\citeauthoryear{Sicat, Li, Choi, Cordeil, Jeong, Bach, and Pfister}{Sicat et~al\mbox{.}}{2018}]%
        {sicat2018dxr}
\bibfield{author}{\bibinfo{person}{Ronell Sicat}, \bibinfo{person}{Jiabao Li}, \bibinfo{person}{Junyoung Choi}, \bibinfo{person}{Maxime Cordeil}, \bibinfo{person}{Won-Ki Jeong}, \bibinfo{person}{Benjamin Bach}, {and} \bibinfo{person}{Hanspeter Pfister}.} \bibinfo{year}{2018}\natexlab{}.
\newblock \showarticletitle{DXR: A toolkit for building immersive data visualizations}.
\newblock \bibinfo{journal}{\emph{IEEE transactions on visualization and computer graphics}} \bibinfo{volume}{25}, \bibinfo{number}{1} (\bibinfo{year}{2018}), \bibinfo{pages}{715--725}.
\newblock


\bibitem[\protect\citeauthoryear{Slater, Brogni, and Steed}{Slater et~al\mbox{.}}{2003}]%
        {slater2003physiological}
\bibfield{author}{\bibinfo{person}{Mel Slater}, \bibinfo{person}{Andrea Brogni}, {and} \bibinfo{person}{Anthony Steed}.} \bibinfo{year}{2003}\natexlab{}.
\newblock \showarticletitle{Physiological responses to breaks in presence: A pilot study}. In \bibinfo{booktitle}{\emph{Presence 2003: The 6th annual international workshop on presence}}, Vol.~\bibinfo{volume}{157}. Citeseer.
\newblock


\bibitem[\protect\citeauthoryear{Slater and Steed}{Slater and Steed}{2000}]%
        {slater2000virtual}
\bibfield{author}{\bibinfo{person}{Mel Slater} {and} \bibinfo{person}{Anthony Steed}.} \bibinfo{year}{2000}\natexlab{}.
\newblock \showarticletitle{A virtual presence counter}.
\newblock \bibinfo{journal}{\emph{Presence}} \bibinfo{volume}{9}, \bibinfo{number}{5} (\bibinfo{year}{2000}), \bibinfo{pages}{413--434}.
\newblock


\bibitem[\protect\citeauthoryear{sqadia.com}{sqadia.com}{2024}]%
        {sqadiaAnatomy}
\bibfield{author}{\bibinfo{person}{sqadia.com}.} \bibinfo{year}{Accessed in 2024}\natexlab{}.
\newblock \bibinfo{title}{Introduction to Anatomy SUBDIVISIONS | Made Easy for Medical Students}.
\newblock
\newblock
\urldef\tempurl%
\url{https://www.youtube.com/watch?v=q6fQf6VLDOY}
\showURL{%
\tempurl}


\bibitem[\protect\citeauthoryear{Subramonyam}{Subramonyam}{2015}]%
        {subramonyam2015sigchi}
\bibfield{author}{\bibinfo{person}{Hariharan Subramonyam}.} \bibinfo{year}{2015}\natexlab{}.
\newblock \showarticletitle{SIGCHI: magic mirror-embodied interactions for the quantified self}. In \bibinfo{booktitle}{\emph{Proceedings of the 33rd Annual ACM Conference Extended Abstracts on Human Factors in Computing Systems}}. \bibinfo{pages}{1699--1704}.
\newblock


\bibitem[\protect\citeauthoryear{Suzuki, Kazi, Wei, DiVerdi, Li, and Leithinger}{Suzuki et~al\mbox{.}}{2020}]%
        {suzuki2020realitysketch}
\bibfield{author}{\bibinfo{person}{Ryo Suzuki}, \bibinfo{person}{Rubaiat~Habib Kazi}, \bibinfo{person}{Li-Yi Wei}, \bibinfo{person}{Stephen DiVerdi}, \bibinfo{person}{Wilmot Li}, {and} \bibinfo{person}{Daniel Leithinger}.} \bibinfo{year}{2020}\natexlab{}.
\newblock \showarticletitle{Realitysketch: Embedding responsive graphics and visualizations in AR through dynamic sketching}. In \bibinfo{booktitle}{\emph{Proceedings of the 33rd Annual ACM Symposium on User Interface Software and Technology}}. \bibinfo{pages}{166--181}.
\newblock


\bibitem[\protect\citeauthoryear{Tohidi, Buxton, Baecker, and Sellen}{Tohidi et~al\mbox{.}}{2006}]%
        {tohidi2006getting}
\bibfield{author}{\bibinfo{person}{Maryam Tohidi}, \bibinfo{person}{William Buxton}, \bibinfo{person}{Ronald Baecker}, {and} \bibinfo{person}{Abigail Sellen}.} \bibinfo{year}{2006}\natexlab{}.
\newblock \showarticletitle{Getting the right design and the design right}. In \bibinfo{booktitle}{\emph{Proceedings of the SIGCHI conference on Human Factors in computing systems}}. \bibinfo{pages}{1243--1252}.
\newblock


\bibitem[\protect\citeauthoryear{Tong, Chen, Xia, Lo, Yuan, Bach, and Qu}{Tong et~al\mbox{.}}{2023}]%
        {tong1912exploring}
\bibfield{author}{\bibinfo{person}{Wai Tong}, \bibinfo{person}{Zhutian Chen}, \bibinfo{person}{Meng Xia}, \bibinfo{person}{Leo Yu-Ho Lo}, \bibinfo{person}{Linping Yuan}, \bibinfo{person}{Benjamin Bach}, {and} \bibinfo{person}{Huamin Qu}.} \bibinfo{year}{2023}\natexlab{}.
\newblock \showarticletitle{Exploring interactions with printed data visualizations in augmented reality}.
\newblock \bibinfo{journal}{\emph{IEEE Transactions on Visualization and Computer Graphics}}  \bibinfo{volume}{29} (\bibinfo{year}{2023}), \bibinfo{pages}{418 -- 428}.
\newblock
Issue 1.


\bibitem[\protect\citeauthoryear{VEGJ}{VEGJ}{2023}]%
        {blocking101}
\bibfield{author}{\bibinfo{person}{DAVOD VEGJ}.} \bibinfo{year}{Last accessed in 2023}\natexlab{}.
\newblock \bibinfo{title}{BLOCKING 101 How directors tell stories with movement}.
\newblock
\newblock
\urldef\tempurl%
\url{https://dramatics.org/blocking-101/}
\showURL{%
\tempurl}


\bibitem[\protect\citeauthoryear{Wang}{Wang}{2020}]%
        {wang2020humor}
\bibfield{author}{\bibinfo{person}{Yunwen Wang}.} \bibinfo{year}{2020}\natexlab{}.
\newblock \showarticletitle{Humor and camera view on mobile short-form video apps influence user experience and technology-adoption intent, an example of TikTok (DouYin)}.
\newblock \bibinfo{journal}{\emph{Computers in human behavior}}  \bibinfo{volume}{110} (\bibinfo{year}{2020}), \bibinfo{pages}{106373}.
\newblock


\bibitem[\protect\citeauthoryear{Wang, Wang, Farinella, Murray-Rust, Henry~Riche, and Bach}{Wang et~al\mbox{.}}{2019}]%
        {wang2019comparing}
\bibfield{author}{\bibinfo{person}{Zezhong Wang}, \bibinfo{person}{Shunming Wang}, \bibinfo{person}{Matteo Farinella}, \bibinfo{person}{Dave Murray-Rust}, \bibinfo{person}{Nathalie Henry~Riche}, {and} \bibinfo{person}{Benjamin Bach}.} \bibinfo{year}{2019}\natexlab{}.
\newblock \showarticletitle{Comparing effectiveness and engagement of data comics and infographics}. In \bibinfo{booktitle}{\emph{Proceedings of the 2019 CHI Conference on Human Factors in Computing Systems}}. \bibinfo{pages}{1--12}.
\newblock


\bibitem[\protect\citeauthoryear{Yang, Kwak, Lee, and Kim}{Yang et~al\mbox{.}}{2023}]%
        {yang2023beyond}
\bibfield{author}{\bibinfo{person}{Saelyne Yang}, \bibinfo{person}{Sangkyung Kwak}, \bibinfo{person}{Juhoon Lee}, {and} \bibinfo{person}{Juho Kim}.} \bibinfo{year}{2023}\natexlab{}.
\newblock \showarticletitle{Beyond Instructions: A Taxonomy of Information Types in How-to Videos}. In \bibinfo{booktitle}{\emph{Proceedings of the 2023 CHI Conference on Human Factors in Computing Systems}}. \bibinfo{pages}{1--21}.
\newblock


\bibitem[\protect\citeauthoryear{Yang, Dwyer, Marriott, Jenny, and Goodwin}{Yang et~al\mbox{.}}{2020}]%
        {yang2020tilt}
\bibfield{author}{\bibinfo{person}{Yalong Yang}, \bibinfo{person}{Tim Dwyer}, \bibinfo{person}{Kim Marriott}, \bibinfo{person}{Bernhard Jenny}, {and} \bibinfo{person}{Sarah Goodwin}.} \bibinfo{year}{2020}\natexlab{}.
\newblock \showarticletitle{Tilt map: Interactive transitions between choropleth map, prism map and bar chart in immersive environments}.
\newblock \bibinfo{journal}{\emph{IEEE Transactions on Visualization and Computer Graphics}} \bibinfo{volume}{27}, \bibinfo{number}{12} (\bibinfo{year}{2020}), \bibinfo{pages}{4507--4519}.
\newblock


\bibitem[\protect\citeauthoryear{Zhou, Huang, and Chan}{Zhou et~al\mbox{.}}{2024a}]%
        {zhou2024epigraphics}
\bibfield{author}{\bibinfo{person}{Tongyu Zhou}, \bibinfo{person}{Jeff Huang}, {and} \bibinfo{person}{Gromit Chan}.} \bibinfo{year}{2024}\natexlab{a}.
\newblock \showarticletitle{Epigraphics: Message-Driven Infographics Authoring}. In \bibinfo{booktitle}{\emph{Proceedings of the 2024 CHI Conference on Human Factors in Computing Systems}} (Honolulu, HI, USA) \emph{(\bibinfo{series}{CHI '24})}.
\newblock
\urldef\tempurl%
\url{https://doi.org/10.1145/3613904.3642172}
\showDOI{\tempurl}


\bibitem[\protect\citeauthoryear{Zhou, Yang, Chan, Chung, and Huang}{Zhou et~al\mbox{.}}{2024b}]%
        {zhou2024portalink}
\bibfield{author}{\bibinfo{person}{Tongyu Zhou}, \bibinfo{person}{Joshua~Kong Yang}, \bibinfo{person}{Vivian~Hsinyueh Chan}, \bibinfo{person}{Ji~Won Chung}, {and} \bibinfo{person}{Jeff Huang}.} \bibinfo{year}{2024}\natexlab{b}.
\newblock \showarticletitle{PortalInk: 2.5D Visual Storytelling with SVG Parallax and Waypoint Transitions}. In \bibinfo{booktitle}{\emph{Proceedings of the 37th Annual ACM Symposium on User Interface Software and Technology}} (Pittsburgh, PA, USA) \emph{(\bibinfo{series}{UIST '24})}. 16.
\newblock
\showISBNx{9798400706288}
\urldef\tempurl%
\url{https://doi.org/10.1145/3654777.3676376}
\showDOI{\tempurl}


\bibitem[\protect\citeauthoryear{Zhu, Xu, Zhang, Chen, and Evans}{Zhu et~al\mbox{.}}{2020}]%
        {zhu2020health}
\bibfield{author}{\bibinfo{person}{Chengyan Zhu}, \bibinfo{person}{Xiaolin Xu}, \bibinfo{person}{Wei Zhang}, \bibinfo{person}{Jianmin Chen}, {and} \bibinfo{person}{Richard Evans}.} \bibinfo{year}{2020}\natexlab{}.
\newblock \showarticletitle{How health communication via Tik Tok makes a difference: A content analysis of Tik Tok accounts run by Chinese provincial health committees}.
\newblock \bibinfo{journal}{\emph{International journal of environmental research and public health}} \bibinfo{volume}{17}, \bibinfo{number}{1} (\bibinfo{year}{2020}), \bibinfo{pages}{192}.
\newblock


\end{thebibliography}

\appendix

\section{How to Create \probes with the Body Object Model}
The Body Object Model (BOM) treats the full-body as a series of nestable anchor points, similar to HTML tags, in which the designer can nest, or bind, visual props to the body. This nested structure serves two purposes. First, it reinforces the idea that the visualization is an element that can be nested, or blended in, with the presenter, effectively blurring the boundaries between the visualization and the presenter. Second, it forces the designer of the visual performance to define the visualizations in relation to the presenter, with a parent-child hierarchy, making the presenter the primary focus even in its language. We implement BOM with Swift and ARKit as they naturally align with the nestable tree structure that we aim to achieve.

The iPhone's TrueDepth and ARKit software detect and segment the presenter's face from the background within 3\,m of the device's front camera. This provides a relative, proxy depth from the presenter's face, enabling us to incorporate three-dimensional form into the visualizations.

Using the self-facing camera, \textsc{ARKit, VNDetectHumanBodyPose3DRequest, VNDetecthumanHandPoseRequest} respectively generate the body anchors, or the face node, body joints, and hand joints. All body anchors are inherited from~\textsc{SCNNode} and the node positions are overlaid on the diagram of the person and hand. Body anchors or the locations that one can bind a~\textsc{VisNode} with using~\textsc{bind}. Each face, body, and hand joint is managed by the \textsc{FacialCueTracker, SkeletonTracker,} and \textsc{HandTracker} respectively. \textsc{HandTracker} also manages \textsc{GestureStates} which then is used to build an \textsc{Action Sequence}. System provides 10 \textsc{GestureStates} and 5 \textsc{ActionSequences}. Using \textsc{Vishandler} the scripts interactions with \textsc{VisNode} and the body.

The prototype is a proof-of-concept tool for experts to design and perform InfoVids. Using this system requires multidisciplinary knowledge of AR, visualization, and performance. Before programming, the designer needs to design a simple narrative for the visualizations. Then, they need to detail how they will stage the interactions between the visualizations and presenter, much like how blocking is done in theater~\cite{blocking101}. Then, after creating the body-vis bindings, all interactions and gestures to trigger a visualization are coded. Based on our nine-month experience with the system, we find that gestures are dictated by the narrative the designer wishes to convey and need to flexibly adapt to suit each unique presentation (Sec.~\ref{sec:lessonslearned}). Thus, at this time, there is no fixed gestural language defined and this is a design choice.

After designing the InfoVid, the smartphone is positioned such that the front-facing camera faces the presenter. The phone screen provides a mirrored view of the presenter with the visualizations such that they can see how they are engaging with the visualizations in real-time, much like AR filters. This setup allows us to evaluate our research questions with InfoVids that have \textit{genuine} interactions from the presenter~\cite{buchenau2000experience}, unlike Wizard-of-Oz styled setups which use post-processing to overlay the augmentations.

\section{Baseline Video Format}
This section outlines how we made the baseline, a 2D videoconferencing style presentation with slides, as comparable as the presentation with \probes. From our external critiques we found there were two major factors to consider to make the baseline.

First, different takes, or renditions of the performance, and the author's biased knowledge of the system generated different performances. Minute changes in facial expressions, enthusiasm, and dialogue changed the perceived narrative. To account for this, we hired an external actor with no prior knowledge of the study to perform in the videos and ensured that both versions of the presentations used the same take of the performance. 

Next, the differences in presenter body movements caused by the added depth were too subtle. If the differences were hard to find by researchers who were actively critiquing, they would be even less discernible to a non-researcher. To make the differences more apparent, we chose a more traditional 2D video-conferencing style presentation as the baseline, referencing the studies of previous interactive visual systems~\cite{davis2023multimodal, lee2013sketchstory}. 

We controlled for the actor's body language, the layout, and the visualizations and its interactive effects with the presenter. The overall process is summarized in \autoref{fig:editing_process}. 

\subsection{Scripting for a Common Body Language}
To incorporate feedback from the critique, we recorded the performance with two iPhone 13 minis simultaneously. As demonstrated in \autoref{fig:editing_process}, this setup provided two videos with the same performance, one \probe and the other with no overlaid visualizations for the baseline. 

\begin{figure*}[!t]
    \centering
\includegraphics[width=\textwidth]{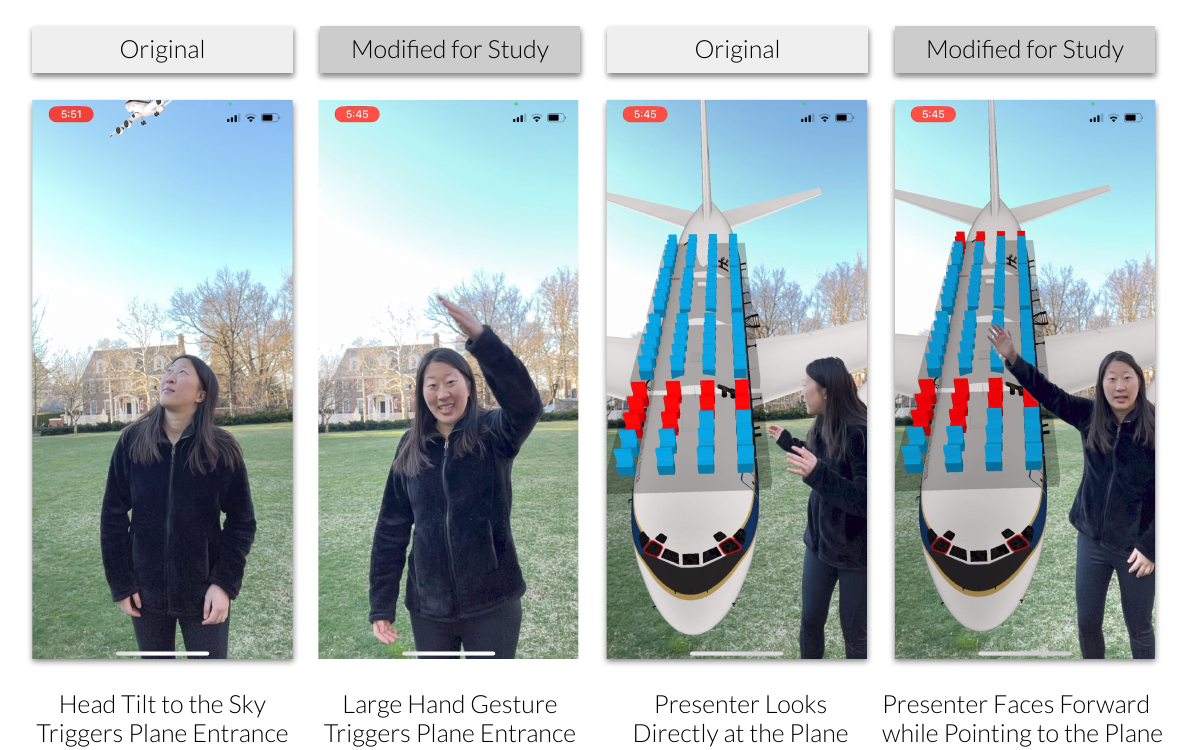}
    \Description{Shows four images side by side. There are four labels on the top, in the order of ‘Original’, ‘Modified for Study’, ‘Original’, ‘Modified for Study’. Each ‘Original’ tag is light gray and each ‘Modified for Study’ tag is gray. For each image underneath the four labels, the performer is seen standing outside on a field. Underneath the first label, ‘Original’, the performer is looking up towards the sky while a 3D plane enters from the top of the image. The caption reads, ‘Head Tilt to the Sky Triggers Plane Entrance’. Underneath the second label, ‘Modified for Study’, the performer is gesturing towards the sky with the left arm, pointing towards the sky. The caption reads, ‘Large Hand Gesture Triggers Plane Entrance’. Under the third label, ‘Original’, the presenter is standing towards the right of an enlarged plane with colored seats, looking at the plane while tilting their body towards the plane. The caption reads, ‘presenter Looks Directly at the Plane’. Under the fourth label, ‘Modified for Study’, the presenter faces forward and uses their right arm to point towards the enlarged plane. The caption reads, ‘Presenter Faces Forward while Pointing to the Plane’.
}
    \caption{Modifications made to original script to preserve common body language for fair comparisons with slides. Presenter looks less immersed in versions modified for the study because they no longer can look directly at the plane visualizations as body language such as looking up to the sky or directly to the plane would contextually make no sense in a videoconferencing presentation with slides---without the visualizations, performer will be staring into the void.}
    \label{fig:cut_examples}
\end{figure*}

The setup to control for confounding factors, however, introduced new problems. Simultaneous recording stipulated a pre-planning process such that the body language could be shared in both presentations. For example, an actor looking directly up towards or at the plane would look awkward and aberrant in a typical slides presentation without the plane, because the actor would be looking into the void (\autoref{fig:cut_examples}). Thus, to minimize such distractions for the viewer, we provided a script that used a common body language  to preserve semantic meaning in both. We instructed the actor to (1) use large, open hand movements with no specific hand or finger gestures and (2) face the camera. As depicted in \autoref{fig:cut_examples}, these modifications detracted from \probe benefits and made the presenter look less immersed, but allowed for fairer comparisons with the baseline.
 
\subsection{Layouts to Preserve Presenter Body Language and Maximize Visualization View}
Because we are interested in the effects of perceived performer body language with the visualizations using \probes, we chose a videoconferencing layout that would clearly capture presenter body language and the visualizations for the baseline. We referenced pre-existing videoconferencing formats in-the-wild and disregarded formats that removed or minimized the size of the presenter to make the visualization the main view. 

While it is possible to stage a visual performance where the full-body is shown, we decided to opt for a version that only showed the upper torso for a few reasons. First, we wanted the baseline to reflect a realistic problem that commonly happens in videoconferencing systems: the cut-off of body language because of the `boxing' out of the presenter. In addition, we needed to use the same take of the performance for the experiment, but some body movements while using \probes were impossible to translate in a 2D setting even with a scripted common body language. For example, \probes encouraged use of full-body movements such as walking back and forth in 3D space or pointing to different parts of the plane, but these movements obstructed the narrative in a 2D videoconferencing setting. However, we kept the upper-half of the torso, so that viewers could still have a view of the performer's hand gestures and facial expressions.  

We also controlled for how much the presenter's body occupied the video frame across different videos. We chose horizontal baseline layout that placed slides to the left of the presenter at a 3:1 ratio~\cite{sqadiaAnatomy} to maximize views of the presenter and the visualization. We also applied \textit{rule of thirds}, a video composition technique often applied in professional videos to create videos that direct the viewer attention to the presenter~\cite{luo2008photo}. As the orange overlay shows (\autoref{fig:editing_process}), we divided the presenter video horizontally and vertically into three parts and placed the presenter's face just above the top vertical line while leaving some space above the head. If the presenter moved around in the video, we zoomed in or translated the video with the presenter to maintain the rule of thirds, using post-hoc video editing. The end effect is comparable to auto zoom-in techniques following a moving presenter in videoconferencing. 

However, the horizontal setup is a trade-off. We exchanged control over phone orientation to maximize screen real estate. Using the vertical orientation of the phone limits the size of the slides and the presenter. Horizontal orientation of the \probe version cuts off half the presenter's body and unfairly disadvantage \probes, which enables full-body performances. On the other hand, as seen in \autoref{fig:editing_process}, the horizontal layout enables a full view of the slides and a sizable view of the presenter and their hand gestures for the baseline. While this setup clips the lower body and some of the presenter's hands at times, these are not atypical formats in video conferencing. Thus, the horizontal layout is a compromise that was necessary for a more equitable comparison between the \probes and the baseline.

\subsection{Post-hoc Synchronization}
In a typical slide presentation, the performer does not have direct control over the visualizations. They can only sync their movements with it. However, because one of the benefits of \probes is the performer's ability to directly control the visualization we synced the animations with the presenter's movements to simulate a well-rehearsed presentation. Using Adobe Premiere Pro, we post-hoc synced the video of the presenter with no virtual overlays with the timings of the animations, gesture triggers, and performer's movements in \probes. To control for effects caused by different visualization props, we also used the same visualization props and animation effects used for \probes, except for \injuryVis where the presenter's body was replaced by a silhouette. These choices dilute \probes advantage to bind visualizations to the body as many presentations in-the-wild don't sync meticulously with the visualizations, but it reduces confounds for the study.

\end{document}